\documentclass[aps,prd,superscriptaddress,groupedaddress,nofootinbib]{article}
\usepackage{dcolumn}
\usepackage{bm}
\usepackage{lscape}
\usepackage{amsmath,amssymb,graphicx}
\usepackage{epsfig}
\usepackage{pstricks}
\usepackage{multirow}
\usepackage[bookmarks,linktocpage]{hyperref}
\usepackage{booktabs}
\usepackage{graphicx}
\usepackage{axodraw4j}
\usepackage{subfigure}
\usepackage{cite}
\usepackage{rotating}
\usepackage{float}
\usepackage{xcolor}
\usepackage{soul} % to cross out text
\usepackage[a4paper, left=2.5cm, right=2.5cm, top=2.5cm, bottom=2.5cm]{geometry}
\usepackage{comment} 
\usepackage{cancel}

\usepackage{array}
\newcolumntype{L}[1]{>{\raggedright\let\newline\\\arraybackslash\hspace{0pt}}m{#1}}
\newcolumntype{C}[1]{>{\centering\let\newline\\\arraybackslash\hspace{0pt}}m{#1}}
\newcolumntype{R}[1]{>{\raggedleft\let\newline\\\arraybackslash\hspace{0pt}}m{#1}}

\RequirePackage{lineno}

\begin{document}

\begin{flushright}
SI-HEP-2018-15
\end{flushright}

\setlength{\arraycolsep}{1.5pt}

\begin{center}
{\Large \textbf{Single production of vector-like quarks with large width\\ at the Large Hadron Collider}}

\vskip.1cm
\end{center}

\vskip0.2cm

\begin{center}
\textbf{{Alexandra Carvalho$^{1}$, Stefano Moretti$^{2,3,4}$, Dermot O'Brien$^{2,3}$,\\ Luca Panizzi$^{5,2}$ and Hugo Prager$^{6,2,3}$} \vskip 8pt }

{\small
$^1$\textit{NICPB, Akadeemia tee, Tallinn, Estonia}\\[0pt]
$^2$\textit{School of Physics and Astronomy, University of Southampton, Highfield, Southampton SO17 1BJ, UK}\\[0pt]
\vspace*{0.1cm} $^3$\textit{Particle Physics Department, Rutherford Appleton Laboratory, Chilton, Didcot, Oxon OX11 0QX, UK}\\[0pt]
\vspace*{0.1cm} $^4$\textit{Physics Department, CERN, CH-1211, Geneva 23, Switzerland}\\[0pt]
\vspace*{0.1cm} $^5$\textit{Dipartimento di Fisica, Universit\`a di Pisa and INFN, Sezione di Pisa, Largo Pontecorvo 3, I-56127 Pisa, Italy}\\[0pt]
\vspace*{0.1cm} $^6$\textit{Theoretische Physik 1, Universit\"at Siegen, Walter-Flex-Stra\ss e 3, D-57068 Siegen, Germany}\\[0pt]
}
\end{center}

\begin{abstract} 
Vector-Like Quarks (VLQs) are predicted by several theoretical scenarios of new physics and, having colour quantum numbers, can copiously be produced at the Large Hadron Collider (LHC), so long that their mass is in the testable kinematic regime of such a machine. While it would be convenient to assume that such objects are narrow and can be treated in the so-called Narrow Width Approximation (NWA), this is not always possible, owing to the fact that couplings and particle content of such new physics scenarios are not generally constrained, so that a large value of the former and/or a large variety of VLQ decay channels into the latter can contribute to generate a large decay width for such extra quarks. We have addressed here the issue of how best to
tackle in LHC analysis the presence of large (and model-dependent) interference effects between different VLQ production and decay channels as well as between these and the corresponding irreducible background. We have confined ourselves to the case of single production of VLQs, which is rapidly becoming a channel of choice in experimental searches owing to the ever increasing limits on their mass, in turn depleting the yield of the historically well-established double production channel. Indeed, this poses a further challenge, as the former is model-dependent while the latter is essentially not. Despite these conditions, we show here that an efficient  approach is possible, which retains to a large extent a degree of model independence in phenomenological studies of such VLQ dynamics at the LHC.
\end{abstract}


\tableofcontents

\newpage

%%%%%%%%%%%%%%%%%%%%%%%%%%%%%%%%%%%%%%%%%%%%%%%%%%%%%%%%%%%%%%%%%%%%%%%%%
%%%%%%%%%%%%%%%%%%%%%%%%%%%%%%%%%%%%%%%%%%%%%%%%%%%%%%%%%%%%%%%%%%%%%%%%%
\section{Introduction}

There is no reason why additional heavy quarks may not exist in Nature:   clearly, though, these ought to be heavier than the top quark. However, following the discovery of a Higgs boson at the Large Hadron Collider (LHC) \cite{Aad:2012tfa,Chatrchyan:2012xdj} closely resembling the one embedded in  the Standard Model (SM), the existence of a fourth generation of chiral quarks (i.e., with SM-like $V-A$ structure in gauge boson charged currents) has been disfavoured \cite{Djouadi:2012ae,Eberhardt:2012gv}\footnote{If the Higgs sector is enlarged with respect to the SM to contain additional  (pseudo)scalar states \cite{Alves:2013dga,Banerjee:2013hxa,Holdom:2014bla}, or if more than one heavy quark multiplet is introduced \cite{Bizot:2015zaa}, new chiral quarks are allowed.}. A possibility for the existence of additional heavy quarks, even in the presence of a SM Higgs sector, is afforded by Vector-Like Quarks
(VLQs), i.e.,  heavy spin 1/2 states that transform as triplets under colour but, differently from the SM quarks, their Left- and Right-Handed (LH and RH) couplings have the same Electro-Weak (EW) quantum numbers. 

VLQs are predicted by several theoretical constructs: e.g., models with a gauged flavour group \cite{Davidson:1987tr,Babu:1989rb,Grinstein:2010ve,Guadagnoli:2011id},  non-minimal
Supersymmetric  scenarios \cite{Moroi:1991mg,Moroi:1992zk,Babu:2008ge,Martin:2009bg,Graham:2009gy,Martin:2010dc}, Grand Unified Theories  \cite{Rosner:1985hx,Robinett:1985dz}), little Higgs  \cite{ArkaniHamed:2002qy,Schmaltz:2005ky} and composite Higgs  \cite{Dobrescu:1997nm,Chivukula:1998wd,He:2001fz,Hill:2002ap,Agashe:2004rs,Contino:2006qr,Barbieri:2007bh,Anastasiou:2009rv}
models, to name but a few. In most of these frameworks VLQs appear as partners of the third generation of quarks, having an EW coupling to top and bottom quarks. 
They are accessible at the LHC detectors in a variety of final states \cite{AguilarSaavedra:2009es,Okada:2012gy,DeSimone:2012fs,Buchkremer:2013bha,Aguilar-Saavedra:2013qpa}, some of these already explored by the ATLAS and CMS collaborations~\cite{Aaboud:2017qpr, Aaboud:2017zfn, Aaboud:2018xuw, Sirunyan:2017usq, Sirunyan:2017pks ,Sirunyan:2018omb,Sirunyan:2017ynj,Sirunyan:2018fjh,ATLAS:2016ovj}. 

Up to now, experimental searches have been designed based  on a simplified scenario for modelling the VLQ dynamics, assuming that only one new VLQ is present beyond the SM and parameterising its production, cross-section ($\sigma$), and decay, Branching Ratio (BR), rates using the Narrow Width Approximation (NWA)~\cite{Aaboud:2017qpr, Aaboud:2017zfn, Aaboud:2018xuw, Sirunyan:2017usq, Sirunyan:2017pks ,Sirunyan:2018omb}, with the exception of~\cite{Sirunyan:2017ynj,Sirunyan:2018fjh} where large width scenarios were also considered. In reality, both aspects need not be true. On the one hand, most of the aforementioned VLQ models predict in general the existence of a new {quark sector}, which implies the presence of more than just one new coloured state. The possibility of reinterpreting experimental data in the context of scenarios with multiple VLQs has been addressed in~\cite{Barducci:2014ila,Barducci:2014gna}. Furthermore, especially when such states are (nearly) degenerate in mass, significant interference effects amongst VLQs may occur that would alter the overall yield \cite{Barducci:2013zaa}. On the other hand, VLQs can have a large width, both in virtue of their mass (much larger than the top quark one) and the variety of decay channels they can undergo, including neutral current ones: contributions from large widths can in turn lead to significant effects, also induced by interferences with the irreducible background, onto the sensitivity of current experimental analyses. Analyses of large width effects in the case of VLQ pair production, i.e., even when the VLQ production stage is due to QCD interactions only, have been performed in~\cite{Moretti:2016gkr,Moretti:2017qby} (see also \cite{Prager:2017owg,Prager:2017hnt}). 

In this paper we  consider singly produced VLQs mixing with either third or light generations of SM quarks. 
Unlike VLQ pair production, wherein the case of mixing with third generation quarks is induced solely by model-independent QCD interactions while the one 
with  light generations also sees a generally subleading (because quark-antiquark induced) EW component, all processes yielding single VLQs in the LHC proceed via model-dependent EW interactions which enter already at the production level connecting the production rate to the aforementioned large width and interference effects.  In order  to interpret the limits of an experimental search on a given range of cross-sections one must determine the size of the VLQ couplings which generate such cross-sections. If the couplings are large, finite width effect and/or interference effects may be not negligible, and it becomes imperative to take them into account in the experimental searches. The exact relation between the EW couplings and those effects is rather model-dependent, resulting on many possibilities for signal modelling.  
{In this paper,} we propose strategies for presenting the results of experimental analyses in a simple, though solid, model-independent framework, allowing a straightforward re-interpretation of experimental data. A similar approach has been already used in its simplest form in Refs.~\cite{Sirunyan:2017ynj,Sirunyan:2018fjh}. Such strategy should limit the need of performing numerical recasting for the reinterpretation of results: the possibility of having ways to avoid numerical recasting (when possible) is becoming pressing as experimental searches rely more and more upon numerical  frameworks which cannot be easily reproduced in tools for phenomenological analyses, such as boosted decision trees or multi-variate analyses. 

The plan of the paper is as follows. In  Section~\ref{sec:parameterisation} we describe the analytical model-independent parameterisation and we discuss the issues induced by interference contributions. In Section~\ref{sec:pheno} we perform the phenomenological analysis: first, in Section~\ref{sec:setup},  we set up our framework, then, in Section~\ref{sec:recast}, we undertake a numerical recasting of a CMS search to obtain bounds for VLQs with large width, finally, in Section~\ref{sec:reinterpretation}, we adopt an analytical parameterisation and compare it to  numerical results, also discussing how to treat interference in a specific benchmark. In the final section we draw our conclusions.

%%%%%%%%%%%%%%%%%%%%%%%%%%%%%%%%%%%%%%%%%%%%%%%%%%%%%%%%%%%%%%%%%%%%%%%%%
%%%%%%%%%%%%%%%%%%%%%%%%%%%%%%%%%%%%%%%%%%%%%%%%%%%%%%%%%%%%%%%%%%%%%%%%%
\section{Model-independent parameterisation of the cross-section} \label{sec:parameterisation}

Apart from the aforementioned small contamination due to EW interactions in the case of mixing with light quarks,  
in processes of \textit{VLQ pair production}, the VLQs are produced through their QCD couplings to the SM gluons and subsequently decay to the SM quarks through couplings induced by their mixing which, together with the mass  of the VLQs, are the free parameters of the Beyond the SM (BSM) physics scenario under consideration. 
{The advantage, from a phenomenological point of view, of pair production processes is thus given by the fact that the production of VLQs is essentially only dependent upon their mass, as the QCD coupling is a SM parameter unmodified by new physics.}

In processes of \textit{VLQ single production}, in contrast, the same type of couplings which induce the VLQ decays also determine the production rate, i.e., the cross-section of single VLQ production processes depends on their couplings to SM quarks, irrespectively of their mixing patterns. The size of such couplings determines for which VLQ mass the cross-section of single production becomes dominant with respect to pair-production at a given collider energy. Moreover, such couplings also contribute to the partial width of the VLQ in their decay channels to SM states. The total width of the VLQ, however, can be increased not only by increasing the couplings of the VLQ with SM quarks and bosons, but also by including further decay channels to other new physics states. In the case of a VLQ with large width, the latter hypothesis is preferable from a phenomenological point of view, as large couplings of VLQs with SM quarks are in general strongly constrained by oblique observables and Higgs couplings~\cite{Chen:2017hak,Moretti:2016gkr}. To perform an analysis of single production of VLQs with large width it is essential to separate the effects which are purely due to the large width from the dependence of the cross-section on the VLQ couplings. As we shall see, in the context of our analysis, such a goal can be achieved considering a suitable parameterisation which factors out the contributions of different couplings.

\begin{figure}[ht!]
\centering
\begin{tabular}{c|c}
\toprule
\multicolumn{2}{c}{quark-boson-boson}\\
\midrule
LW and NWA & only LW \\
\begin{picture}(90,40)(0,-10)
  \SetWidth{1.0}
  \SetColor{Black}
  \Gluon(10,10)(30,10){2}{4}
  \Text(8,10)[rc]{\small{$g$}}
  \Line(10,30)(30,30)
  \Text(8,30)[rc]{\small{$q$}}
  \SetWidth{2.0}
  \SetColor{Red}
  \Line(30,10)(30,30)
  \Text(32,20)[lc]{\small{\Red{$Q$}}}
  \Line(30,10)(50,10)
  \Text(40,7)[ct]{\small{\Red{$Q$}}}
  \SetWidth{1.0}
  \SetColor{Black}
  \Photon(30,30)(50,30){2}{4}
  \Line[dash](30,30)(50,30)
  \Text(52,30)[lc]{\small{$V,H$}}
  \Line(50,10)(70,20)
  \Text(72,20)[lc]{\small{$q$}}
  \Photon(50,10)(70,0){2}{4}
  \Line[dash](50,10)(70,0)
  \Text(72,0)[lc]{\small{$V,H$}}
\end{picture}
\begin{picture}(110,40)(0,-10)
  \SetWidth{1.0}
  \SetColor{Black}
  \Gluon(10,10)(30,20){2}{4}
  \Text(8,10)[rc]{\small{$g$}}
  \Line(10,30)(30,20)
  \Text(8,30)[rc]{\small{$q$}}
  \Line(30,20)(50,20)
  \Text(40,17)[ct]{\small{$q$}}
  \Photon(50,20)(70,30){2}{4}
  \Line[dash](50,20)(70,30)
  \Text(72,30)[lc]{\small{$V,H$}}
  \SetWidth{2.0}
  \SetColor{Red}
  \Line(50,20)(70,10)
  \Text(57,13)[ct]{\small{\Red{$Q$}}}
  \SetWidth{1.0}
  \SetColor{Black}
  \Photon(70,10)(90,20){2}{4}
  \Line[dash](70,10)(90,20)
  \Text(92,20)[lc]{\small{$V,H$}}
  \Line(70,10)(90,0)
  \Text(92,0)[lc]{\small{$q$}}
\end{picture}  
& 
\begin{picture}(70,60)(0,0)
  \SetWidth{1.0}
  \SetColor{Black}
  \Gluon(10,10)(30,10){2}{4}
  \Text(8,10)[rc]{\small{$g$}}
  \Line(10,50)(30,50)
  \Text(8,50)[rc]{\small{$q$}}
  \Line(30,10)(30,30)
  \Text(32,20)[lc]{\small{$q$}}
  \SetWidth{2.0}
  \SetColor{Red}
  \Line(30,30)(30,50)
  \Text(32,40)[lc]{\small{\Red{$Q$}}}
  \SetWidth{1.0}
  \SetColor{Black}
  \Photon(30,30)(50,30){2}{4}
  \Line[dash](30,30)(50,30)
  \Text(52,30)[lc]{\small{$V,H$}}
  \Photon(30,50)(50,50){2}{4}
  \Line[dash](30,50)(50,50)
  \Text(52,50)[lc]{\small{$V,H$}}
  \Line(30,10)(50,10)
  \Text(52,10)[lc]{\small{$q$}}
\end{picture}
\\
\midrule
\midrule
\multicolumn{2}{c}{quark-boson-quark}\\
\midrule
LW and NWA & only LW \\
\begin{picture}(90,40)(0,-10)
  \SetWidth{1.0}
  \SetColor{Black}
  \Line(10,10)(30,10)
  \Text(8,10)[rc]{\small{$q$}}
  \Line(10,30)(30,30)
  \Text(8,30)[rc]{\small{$q$}}
  \Photon(30,10)(30,30){2}{4}
  \Line[dash](30,10)(30,30)
  \Text(33,20)[lc]{\small{$V,H$}}
  \SetWidth{2.0}
  \SetColor{Red}
  \Line(30,10)(50,10)
  \Text(40,7)[ct]{\small{\Red{$Q$}}}
  \SetWidth{1.0}
  \SetColor{Black}
  \Line(30,30)(50,30)
  \Text(52,30)[lc]{\small{$q$}}
  \Line(50,10)(70,20)
  \Text(72,20)[lc]{\small{$q$}}
  \Photon(50,10)(70,0){2}{4}
  \Line[dash](50,10)(70,0)
  \Text(72,0)[lc]{\small{$V,H$}}
\end{picture}
\begin{picture}(110,40)(0,-10)
  \SetWidth{1.0}
  \SetColor{Black}
  \Line(10,10)(30,20)
  \Text(8,10)[rc]{\small{$q$}}
  \Line(10,30)(30,20)
  \Text(8,30)[rc]{\small{$q$}}
  \Photon(30,20)(50,20){2}{4}
  \Line[dash](30,20)(50,20)
  \Text(40,16)[ct]{\small{$V,H$}}
  \Line(50,20)(70,30)
  \Text(72,30)[lc]{\small{$q$}}
  \SetWidth{2.0}
  \SetColor{Red}
  \Line(50,20)(70,10)
  \Text(57,13)[ct]{\small{\Red{$Q$}}}
  \SetWidth{1.0}
  \SetColor{Black}
  \Photon(70,10)(90,20){2}{4}
  \Line[dash](70,10)(90,20)
  \Text(92,20)[lc]{\small{$V,H$}}
  \Line(70,10)(90,0)
  \Text(92,0)[lc]{\small{$q$}}
\end{picture}  
&
\begin{picture}(70,60)(0,0)
  \SetWidth{1.0}
  \SetColor{Black}
  \Line(10,10)(30,10)
  \Text(8,10)[rc]{\small{$q$}}
  \Line(10,50)(30,50)
  \Text(8,50)[rc]{\small{$q$}}
  \Photon(30,10)(30,30){2}{4}
  \Line[dash](30,10)(30,30)
  \Text(33,20)[lc]{\small{$V,H$}}
  \SetWidth{2.0}
  \SetColor{Red}
  \Line(30,30)(30,50)
  \Text(32,40)[lc]{\small{\Red{$Q$}}}
  \SetWidth{1.0}
  \SetColor{Black}
  \Line(30,30)(50,30)
  \Text(52,30)[lc]{\small{$q$}}
  \Photon(30,50)(50,50){2}{4}
  \Line[dash](30,50)(50,50)
  \Text(52,50)[lc]{\small{$V,H$}}
  \Line(30,10)(50,10)
  \Text(52,10)[lc]{\small{$q$}}
\end{picture}
\begin{picture}(90,40)(0,-10)
  \SetWidth{1.0}
  \SetColor{Black}
  \Line(10,10)(30,10)
  \Text(8,10)[rc]{\small{$q$}}
  \Line(10,30)(30,30)
  \Text(8,30)[rc]{\small{$q$}}
  \SetWidth{2.0}
  \SetColor{Red}
  \Line(30,10)(30,30)
  \Text(33,20)[lc]{\small{\Red{$Q$}}}
  \SetWidth{1.0}
  \SetColor{Black}
  \Photon(30,30)(50,30){2}{4}
  \Line[dash](30,30)(50,30)
  \Text(52,30)[lc]{\small{$V,H$}}
  \Photon(30,10)(50,10){2}{4}
  \Line[dash](30,10)(50,10)
  \Text(40,6)[ct]{\small{$V,H$}}
  \Line(50,10)(70,20)
  \Text(72,20)[lc]{\small{$q$}}
  \Line(50,10)(70,0)
  \Text(72,0)[lc]{\small{$q$}}
\end{picture}
\\
\bottomrule
\end{tabular}
\caption{\label{fig:topologies} Complete list of the generic topologies for final states compatible with single VLQ production in the 5FS. Topologies on the left column can be described in both the NWA and LW regimes while topologies on the right column are neglected in the NWA approximation. Here, $V$ represents the $W$ and $Z$ bosons of the SM.}
\end{figure}
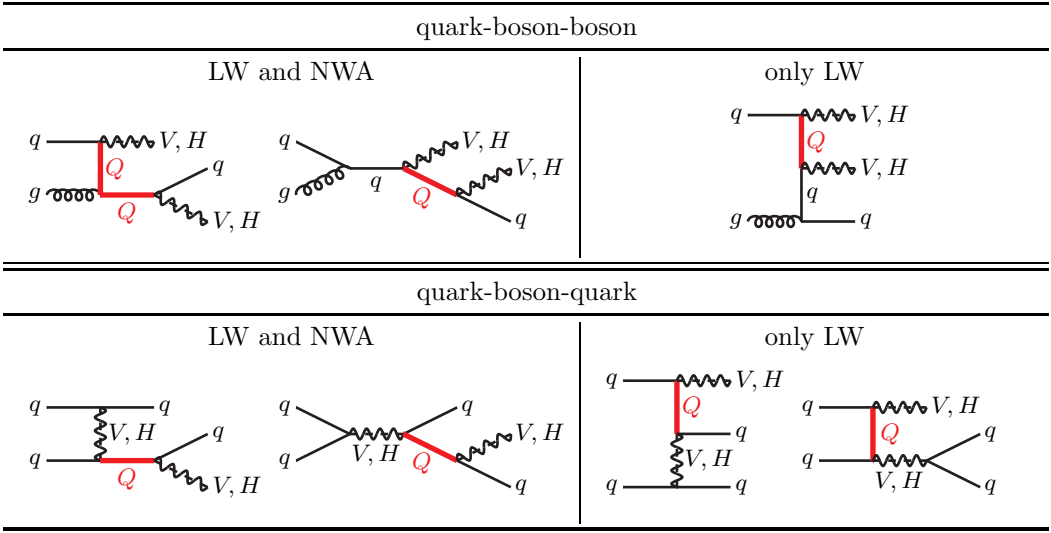

But let us now proceed to define the constituent processes of VLQ production in single mode. In essence, herein,  
%%%In the usual NWA, the on-shellness condition is explicitly imposed in the Monte Carlo (MC) generation. To consider the signal with finite width, instead, 
we consider any topology leading to a  three-particle final state made up by SM objects only  
which contains at least one VLQ propagator. A subtlety, however, arises when deciding upon the initial state.
In fact, 
single production of VLQs can be studied either by considering the proton to be constituted of five quark flavours, the so-called 5-Flavour Scheme (5FS), or else assuming that bottom quarks are not in the initial state but can only appear from gluon splitting, { i.e.} the 4-Flavour Scheme (4FS). Any phenomenological difference between 4FS and 5FS would vanish if it was possible to describe the processes at all orders in QCD perturbation theory\footnote{For specific details we refer to the wide literature on this subject (see for example~\cite{Campbell:2009ss,Frederix:2012dh,Bothmann:2017jfv} for single top production).}. In our analysis we will work in the 5FS, for which the topologies are  represented in Fig.~\ref{fig:topologies}. (Notice that some of these only exist in the Large Width (LW) regime, not in the NWA.)
This choice is simply dictated by the preliminary results of Ref.~\cite{Brooijmans:2018xbu} (Contribution 3), which show that, for a specific process and in the NWA, when kinematical distributions of final state objects computed at LO in the 5FS exhibit sizeable differences with respect to those computed at LO in the 4FS, the distributions obtained through a NLO description of the process are closer in shape to the LO results in the 5FS.

The total cross-section ($\sigma$) can be parameterised by factorising the BSM couplings considering a \textit{reduced} quantity, labelled as $\hat\sigma$, analogously to what was done in~\cite{Buchkremer:2013bha}. In the absence of any kind of interference between different topologies, the model-independent $\hat\sigma$ depends only on the mass and total width of the VLQ. The factorisation of the signal cross-section (labelled with the subscript 
$S$ in the following) can  be written as (see Fig.~\ref{fig:topologies}): 
\begin{equation}
\label{eq:FW}
\sigma_S(C_1,C_2,M_Q,\Gamma_Q)=C_1^2~C_2^2~\hat\sigma_S(M_Q,\Gamma_Q),
\end{equation}
where $C_1$ and $C_2$ are the couplings corresponding to the interactions at both sides of the $Q$ propagator, and $\Gamma_Q = \Gamma(C_i,M_Q,m_{\rm decays})$ is the total width of the VLQ, which depends on its mass, the masses of all its decay products and the couplings through which the VLQ interacts with all the particles it can decay to, including (but not exclusively) $C_1$ and $C_2$. \\

\begin{figure}[ht!]
\centering
\begin{tabular}{cc|cc}
\begin{picture}(90,40)(0,-15)
  \SetWidth{1.0}
  \SetColor{Black}
  \Line(10,10)(30,10)
  \Text(8,10)[rc]{\small{$b$}}
  \Line(10,30)(30,30)
  \Text(8,30)[rc]{\small{$\bar b$}}
  \Photon(30,10)(30,30){2}{4}
  \Text(33,20)[lc]{\small{$W^+$}}
  \SetWidth{2.0}
  \SetColor{Red}
  \Line(30,10)(50,10)
  \Text(40,7)[ct]{\small{\Red{$T$}}}
  \SetWidth{1.0}
  \SetColor{Black}
  \Line(30,30)(50,30)
  \Text(52,30)[lc]{\small{$\bar t$}}
  \Line(50,10)(70,20)
  \Text(72,20)[lc]{\small{$b$}}
  \Photon(50,10)(70,0){2}{4}
  \Text(72,0)[lc]{\small{$W^+$}}
\end{picture}
&
\begin{picture}(70,60)(0,0)
  \SetWidth{1.0}
  \SetColor{Black}
  \Line(10,10)(30,10)
  \Text(8,10)[rc]{\small{$b$}}
  \Line(10,50)(30,50)
  \Text(8,50)[rc]{\small{$\bar b$}}
  \Photon(30,10)(30,30){2}{4}
  \Text(33,20)[lc]{\small{$Z$}}
  \SetWidth{2.0}
  \SetColor{Red}
  \Line(30,30)(30,50)
  \Text(32,40)[lc]{\small{\Red{$T$}}}
  \SetWidth{1.0}
  \SetColor{Black}
  \Line(30,30)(50,30)
  \Text(52,30)[lc]{\small{$\bar t$}}
  \Photon(30,50)(50,50){2}{4}
  \Text(52,50)[lc]{\small{$W^+$}}
  \Line(30,10)(50,10)
  \Text(52,10)[lc]{\small{$b$}}
\end{picture}
&
\begin{picture}(90,40)(0,-15)
  \SetWidth{1.0}
  \SetColor{Black}
  \Line(10,10)(30,10)
  \Text(8,10)[rc]{\small{$b$}}
  \Line(10,30)(30,30)
  \Text(8,30)[rc]{\small{$\bar b$}}
  \Photon(30,10)(30,30){2}{4}
  \Text(33,20)[lc]{\small{$W^+$}}
  \Line(30,10)(50,10)
  \Text(40,7)[ct]{\small{$t$}}
  \Line(30,30)(50,30)
  \Text(52,30)[lc]{\small{$\bar t$}}
  \Line(50,10)(70,20)
  \Text(72,20)[lc]{\small{$b$}}
  \Photon(50,10)(70,0){2}{4}
  \Text(72,0)[lc]{\small{$W^+$}}
\end{picture}
&
\begin{picture}(70,60)(0,0)
  \SetWidth{1.0}
  \SetColor{Black}
  \Line(10,10)(30,10)
  \Text(8,10)[rc]{\small{$b$}}
  \Line(10,50)(30,50)
  \Text(8,50)[rc]{\small{$\bar b$}}
  \Photon(30,10)(30,30){2}{4}
  \Text(33,20)[lc]{\small{$Z$}}
  \Line(30,30)(30,50)
  \Text(32,40)[lc]{\small{$t$}}
  \Line(30,30)(50,30)
  \Text(52,30)[lc]{\small{$\bar t$}}
  \Photon(30,50)(50,50){2}{4}
  \Text(52,50)[lc]{\small{$W^+$}}
  \Line(30,10)(50,10)
  \Text(52,10)[lc]{\small{$b$}}
\end{picture}
\end{tabular}
\caption{\label{fig:topologiesinterference} Example on the left of a subset of interfering topologies in the process $b b \to b W^+ \bar t$ mediated by a $T$ VLQ and different SM bosons. The diagrams on the right show analogous contributions from the SM background, which interferes with the signal.}
\end{figure}
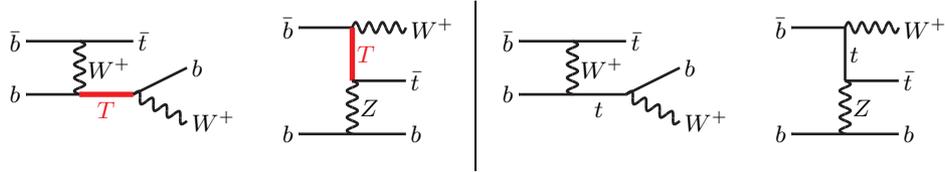

This parameterisation cannot be generalised if interference contributions amongst signal components that contain different coupling content and/or with SM background are non-negligible. In this case the cross-sections for the pure signal and its interference contributions with the irreducible background are written as $\sigma_S$ and $\sigma_{SB}^{\rm{int}}$, where the second represents the interferences amongst signal  ($S$) topologies or with the SM background ($B$). In Fig.~\ref{fig:topologiesinterference} it is shown an example for the $Wtb$ final state: in this case $\sigma_S = \sigma_W + \sigma_Z + \sigma_{WZ}^{\rm{int}}$ and $\sigma_{SB}^{\rm{int}} = \sigma_{WB}^{\rm{int}} + \sigma_{ZB}^{\rm{int}}$ where the subscript of each term represents the bosons propagating in the corresponding topologies. The individual terms of the sum are not necessarily gauge invariant, while the \textit{observable} quantities $\sigma_S$ and $\sigma_{SB}^{\rm{int}}$ are. 

The only possible factorisation in case interference terms are not negligible is:
\begin{eqnarray}
\label{eq:FWinterference}
\sigma_S(C_{1\dots},C_2,M_Q,\Gamma_Q,\chi_Q) &=& C_2^2~\hat\sigma_S(C_{1\dots},M_Q,\Gamma_Q,\chi_Q) \quad\rm{and}\quad \sigma_{SB}^{\rm{int}}=C_2~\hat\sigma^\text{int}_{SB}(C_{1\dots},M_Q,\Gamma_Q,\chi_Q) ,
\end{eqnarray}
where $C_2$ is the coupling corresponding to the interaction between $Q$ and the boson in the final state, $C_{1\dots}$ are the couplings through which $Q$ interacts with the {\it different}  virtual bosons and $\chi_Q$ the dominant chirality of the VLQ couplings, which can be LH or RH depending on the VLQ representation and for total rates only plays a role in interference terms. As an example, a $T$ singlet will decay through charged current to a bottom quark with a dominantly LH chirality~\cite{Buchkremer:2013bha,Aguilar-Saavedra:2013qpa} whereas a $T$ as part of a doublet will produce a dominantly RH bottom: the interference with the SM background, for which the bottom quark can only be produced with LH chirality through charged currents, will be different in the two scenarios.

In the limit of negligible interference between different signal topologies, the pure signal contribution can be approximated as:
\begin{equation}
\sigma_S(C_{1\dots},C_2,M_Q,\Gamma_Q) \sim C_2^2~\sum_{i} C_{1i}^2~\hat\sigma_{Si}(M_Q,\Gamma_Q),
\end{equation}
where the potentially gauge-dependent contribution is assumed to be negligible as well. 

A further limit is represented by scenarios in which the VLQ interacts exclusively with one SM boson. In this case, besides recovering Eq.~\eqref{eq:FW} for the cross-section of the pure signal contribution, the interference with SM background can be written as:
\begin{equation}
\label{eq:FWSMintlimit}
\sigma_{SB}^{\rm{int}}(C_2,M_Q,\Gamma_Q,\chi_Q)=C_2^2~\hat\sigma^\text{int}_{SB}(M_Q,\Gamma_Q,\chi_Q).
\end{equation}

Eqs.~\eqref{eq:FW} to \eqref{eq:FWSMintlimit} are valid in all width regimes, though their range of applicability is different. The values of $\hat\sigma_S$ in the parameterisation of Eq.~\eqref{eq:FW} can be computed for any possible final state and can be used for a model-independent reinterpretation of the results, under the assumption that the model for which the reinterpretation is made does not contain sizeable interference terms of any kind for the considered final state. Such parameterisation, when associated with a proper way to interpret experimental results, discussed below, represents a robust model-independent framework for testing any theoretical scenario predicting VLQs with small or large width against experimental limits in a given channel. This approach has been followed in two CMS analyses~\cite{Sirunyan:2017ynj,Sirunyan:2018fjh}, where $\hat\sigma$  was used for an interpretation of experimental results for single production of VLQs with large width. 
The values of $\hat\sigma_S$ and $\hat\sigma^\text{int}_{SB}$ in Eqs.~\eqref{eq:FWinterference} to \eqref{eq:FWSMintlimit}, however, contain a further model-dependency, represented by the impossibility to factorise the \{$C_{1\dots}$\} couplings in a gauge-invariant way. If interference terms are sizeable, this is an unavoidable consequence and the analysis would have in any case a limited reinterpretation applicability.

%The first has been followed in two CMS analyses~\cite{Sirunyan:2017ynj,Sirunyan:2018fjh} where $\hat\sigma$  was used for an interpretation of experimental results for single production of VLQs with large width.

Both parameterisations can be easily generalised to compute cross-sections at NLO in QCD, as the factorisation of the EW couplings to derive the $\hat\sigma$'s is not affected by the contribution of radiative QCD effects. A NLO calculation of single production of VLQs with finite width is however beyond the scope of this paper.

%%%%%%%%%%%%%%%%%%%%%%%%%%%%%%%%%%%%%%%%%%%%%%%%%%%%%%%%%%%%%%%%%%%%%%%%%
%%%%%%%%%%%%%%%%%%%%%%%%%%%%%%%%%%%%%%%%%%%%%%%%%%%%%%%%%%%%%%%%%%%%%%%%%
\section{Phenomenological analysis}
\label{sec:pheno}

In this section we will provide case studies for the reinterpretation of experimental results of searches for single production of VLQs with finite width, comparing the bounds obtained through a numerical recasting with the results obtained applying the analysis strategies of Sect.~\ref{sec:parameterisation}.

%%%%%%%%%%%%%%%%%%%%%%%%%%%%%%%%%%%%%%%%%%%%%%%%%%%%%%%%%%%%%%%%%%%%%%%%%
\subsection{Benchmarks}
\label{sec:setup}

Analogously to what has already been done in the CMS analyses~\cite{Sirunyan:2017ynj,Sirunyan:2018fjh}, we will consider specific and representative scenarios for the interactions of VLQs with the SM states. Couplings between VLQs and SM quarks and bosons are parameterised as:
\begin{equation}
c_Z = \frac{e}{2 c_w s_w}\kappa_Z, \quad c_W = \frac{e}{\sqrt{2} s_w}\kappa_W \quad\text{and}\quad c_H = \frac{M_Q}{v}\kappa_H,
\label{eq:couplings}
\end{equation}
where $v=246$~GeV is the Higgs Vacuum Expectation Value (VEV), $c_w$ and $s_w$ are the cosine and sine of the weak angle $\theta_w$ and $\kappa_{W,Z,H}$ are coupling strengths. In the limit where only one VLQ representation is present in Nature and the particles are narrow the $\kappa$ parameters accurately approximate the sine of the mixing angle between VLQ and SM quark (for a relation between couplings and mixing angles see, {e.g.} Ref.~\cite{Chen:2017hak}). In the NWA, if the VLQ is a singlet, the coupling parameters are related as $\kappa_Z\simeq\kappa_H\simeq\kappa_W\simeq\kappa$ while, if the VLQ is within a doublet (and assuming that the Yukawa coupling of the other element of the doublet is zero), the coupling parameters become $\kappa_Z\simeq\kappa_H\simeq\kappa$ and $\kappa_W\simeq0$. These relations are not satisfied in the large width regime.

Minimal extensions of the SM with one unique VLQ representation where the VLQ has large couplings are excluded by other constraints (such as EW precision tests or corrections to SM couplings)~\cite{Chen:2017hak,Moretti:2016gkr}. In our case study the large width regime will be achieved by considering the total width as a free parameter but imposing for VLQ singlets and doublets identical relations between the $\kappa$ parameters as in the NWA. We also consider more extreme scenarios where we assume 100\% couplings of the VLQ with each of the SM bosons. 
A summary of the benchmarks we consider is reported in Tab.~\ref{tab:BPs}. Through this procedure the assumption of a  minimal model is relaxed: new physics, in the form of mixings either in the quark or bosonic sectors due to additional representations of VLQs or new gauge bosons or scalar fields, must contribute to generate the relations between couplings.\\
\begin{table}[ht!]
\centering
\begin{tabular}{ccc|c|c}
\toprule
100\% $W$                  & 100\% $Z$                    & 100\% $H$                   & Singlet-like                       & Doublet-like \\
\midrule
$\kappa_Z^q=\kappa_H^q=0$ & $\kappa_W^q=\kappa_H^q=0$ & $\kappa_W^q=\kappa_Z^q=0$ & $\kappa_W^q=\kappa_Z^q=\kappa_H^q$ & $\kappa_W^q=0 $ and $\kappa_Z^q=\kappa_H^q$\\
\bottomrule
\end{tabular}
\caption{\label{tab:BPs} Benchmark points and corresponding coupling relations.}
\end{table}

The event generation has been performed through {\sc MadGraph5\_aMC@NLO}~\cite{Alwall:2011uj,Alwall:2014hca} using the NNPDF3.0 PDF set \cite{Ball:2014uwa} at LO with $\alpha_S$ = 0.118 and setting the QCD renormalisation and PDF factorisation scales to the mass of the VLQ. The generated events have been subsequently processed through {\sc Pythia}\,8~\cite{Sjostrand:2007gs,Sjostrand:2006za} to include  parton showering and hadronisation. Finally, the hadronised events have been analysed through {\sc MadAnalysis}\,5~\cite{Conte:2014zja}, which uses {\sc Delphes\,3}~\cite{deFavereau:2013fsa} for the emulation of detector effects. The VLQ model used for the simulation was implemented in {\sc Feynrules}~\cite{Alloul:2013bka} to obtain the {\tt UFO}~\cite{Degrande:2011ua} model format to be used inside {\sc MadGraph5\_aMC@NLO}. This model is a slightly modified version of the public model described in~\cite{Fuks:2016ftf} where each coupling of the VLQ has been assigned a different label in order to isolate individual interference contributions through the {\sc MadGraph5\_aMC@NLO} syntax {\sc coupling\_order\_i$^\wedge$2==N} in the process of diagram generation, where the {\sc i}'th coupling order is counted {\sc N} times in the squared amplitudes, making it possible to generate separately each term of Eq.~\eqref{eq:FWinterference}.

\begin{figure}[ht!]
\centering
\epsfig{file=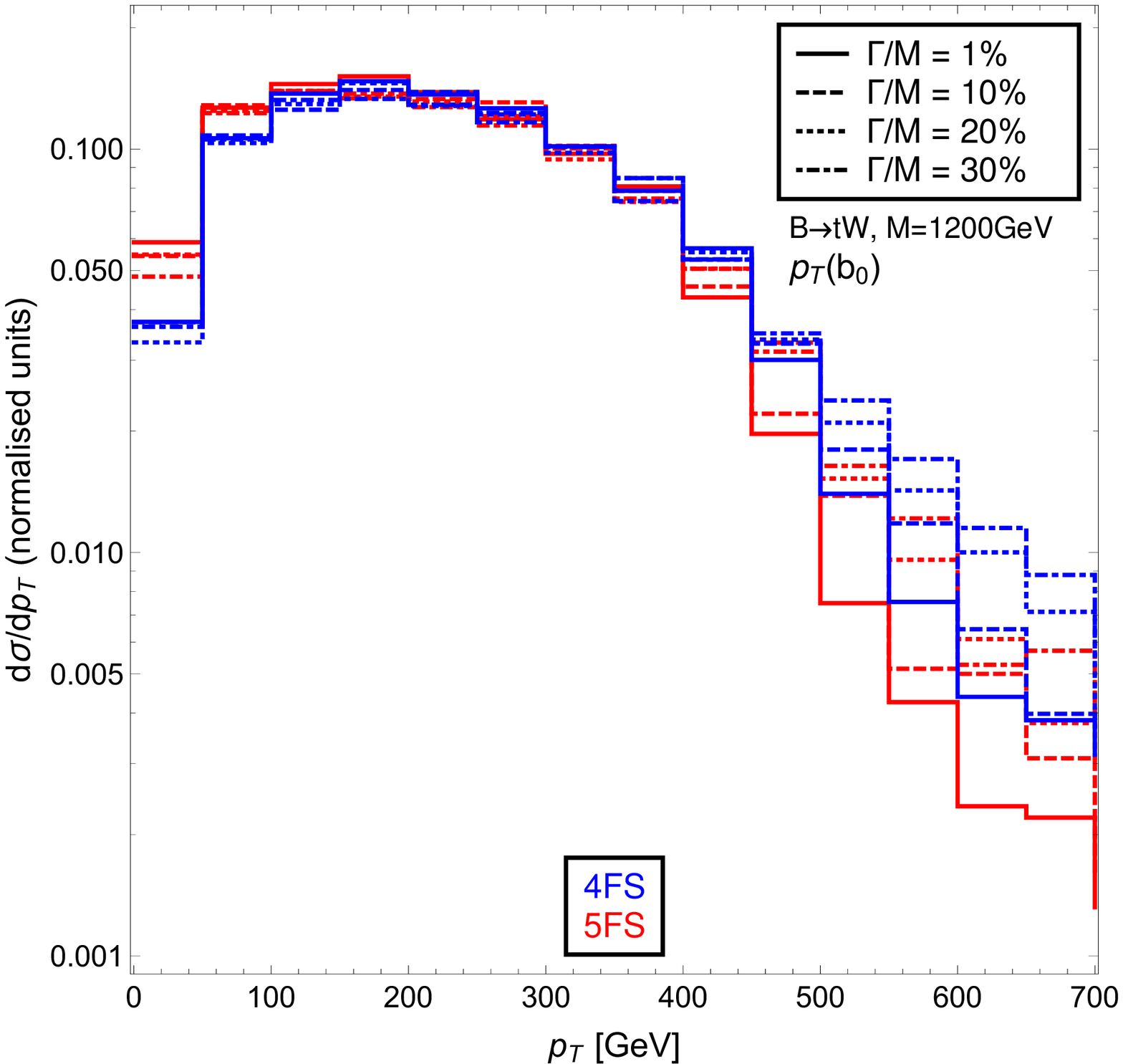 ,width=.45\textwidth}
\epsfig{file=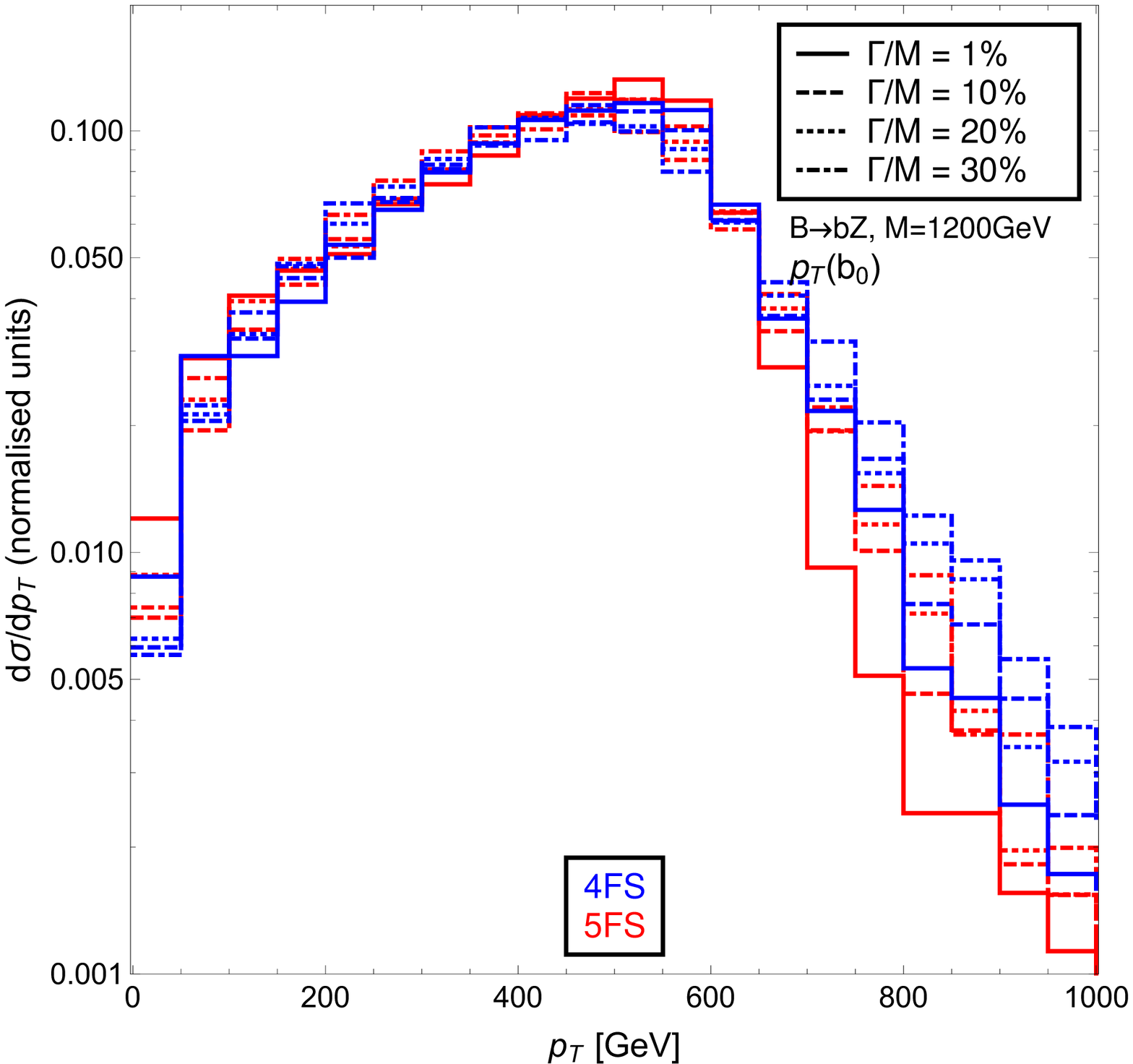 ,width=.45\textwidth}
\caption{\label{fig:distributions} Distributions of the transverse momentum of the leading $b$-tagged jet for two processes. Left panel: $pp\to t W j$ (in the 5FS) or $pp\to tWjb$ (in the 4FS); right panel: $pp\to b Z j$ (in the 5FS) or $pp\to bZjb$ (in the 4FS), with subsequent on-shell decays of the SM states in both cases. The comparison is made for a $B$ mass of 1200 GeV and for different values of the width-over-mass ratio ($\Gamma_Q/M_Q$) of the VLQ.}
\end{figure}

It is not within the scope of this paper to investigate the  impact on  searches due to the choice of the proton scheme. However, as most of the experimental searches performed so far adopt the 4FS (including the one recast in this analysis)  a more quantitative discussion about the choice of the flavour number scheme is in order. As an example, in Fig.~\ref{fig:distributions}, we show the $p_T$ distribution of the leading $b$-jet comparing the 4FS and 5FS at LO for a $B$ VLQ mixing with third generation, in different width regimes, for the processes $pp\to t W j$ and $pp\to b Z j$. The dependence of the distributions on the width of the VLQ can be sizeable, thus potentially affecting the efficiency of experimental cuts. The final state $b$-quark is dominantly generated in the $B$ decay and therefore is fairly central in both 5FS and 4FS in any width regime. The width-dependent differences in the shapes are generally of similar size in both 4FS and 5FS.  

The signal cross-sections are generally of similar size between 4FS and 5FS in the NWA at LO~\cite{Brooijmans:2018xbu}. We had also checked that in all parameter space we probe the cross-sections calculated in the 4FS are always  smaller than the ones calculated on the 5FS. Therefore, the interpretation of results in terms of number of events in the 4FS acts as a lower bound to the a more realistic exclusion bound.

%%%%%%%%%%%%%%%%%%%%%%%%%%%%%%%%%%%%%%%%%%%%%%%%%%%%%%%%%%%%%%%%%%%%%%%%%
\subsection{Recasting and reinterpretation of experimental data}
\label{sec:recast}

As a representative analysis for the recast of experimental data, the CMS search CMS-B2G-16-006~\cite{Sirunyan:2017tfc} has been chosen. This analysis specifically targets a heavy VLQ with charge $+2/3$ or $-4/3$ decaying into a $b$-quark and a $W$ boson, which is produced singly in association with a light flavour quark and a $b$-quark. Thus, in the 4FS, the considered process is $p p \rightarrow T(\to W \, b) \, \bar b \, j$. The search selects events with one lepton (electron or muon in two different signal regions) and at least two jets, one central ($b$-tagged) and one forward. A set of kinematical cuts are also imposed, and we refer to~\cite{Sirunyan:2017tfc} for all the relevant details. In table~\ref{tab:validation}  we compare the efficiencies obtained by our recast  implementation with the ones reported by~\cite{Sirunyan:2017tfc} under the same signal assumptions of the experimental search.    
We stress that, for the purposes of this study, we are not interested in an overly precise reconstruction of the experimental results: our intent is to achieve a realistic analysis framework to demonstrate how to reinterpret experimental data in scenarios where the VLQs have a large width.

\begin{table}[ht!]
\centering
\begin{tabular}{c|ccc}
\hline 
Channel & Electron & Muon & Total \\ 
\hline 
Experimental efficiencies & 1.3\% & 1.4\% & 2.7\% \\ 
\hline 
Recasting efficiencies & 0.98\% & 1.31\% & 2.29\% \\ 
\hline 
\end{tabular} 
\caption{\label{tab:validation} Comparison of the experimental and recasted efficiencies for the different signal regions for a $pp\to T(\to Wb) jb$ signal in the 4FS with $m_T=1000$ GeV in the NWA.}
\end{table}

\begin{figure}[ht!]
\centering
\epsfig{file=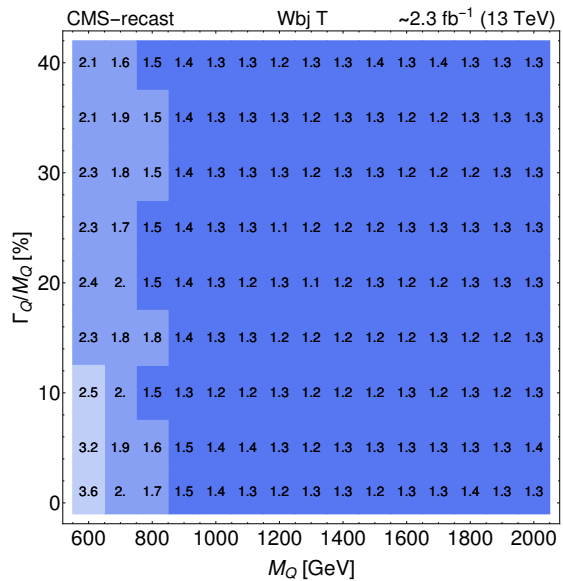,width=.45\textwidth}
\caption{\label{fig:sigma95} 95\% CL largest allowed cross-sections (in pb) for the $Wbj$ final state as a function of $M_T$ and of $\Gamma_T / M_T$ for a VLQ $T$ coupling to the third SM quark generation and decaying 100\% to $bW$.}
\end{figure}

We recast this analysis for the large width case by scanning the  $ \Gamma_Q / M_Q$ from, e.g.,  1\% to 40\% in steps of 5\% and $M_Q$ from 600 to 2000~GeV in steps of 100~GeV. Four processes will be considered: the original target of the search (a third generation $T$ partner in the $p p \rightarrow W \, b \, q$ channel) but also a third generation $B$ partner in the $p p \rightarrow W \, t \, t$ channel, and a light generation $T$ and $B$ partner on the $p p \rightarrow W \, \bar q \, q$ channel (notice that, as we consider the 5FS, a $q$ may also be a $b$).  For the sake of linearity, we first show the results of the recast in the  $p p \rightarrow W \, b \, q$ channel. The values of the the largest allowed observed cross-section at 95\% Confidence Level (CL) on the $(M_Q, \Gamma_Q / M_Q)$ plane that we obtain for this channel are shown in Fig.~\ref{fig:sigma95}. 

% \begin{figure*}[h]\begin{center}
% \epsfig{file=./example,width=.48\textwidth}
% % \includegraphics[width=0.48\textwidth, angle =0 ]{example.pdf}
% \caption{\small{Example for superimposing colormap to values, this is a toy plot made from a table of numbers obtainex with F(x,y) = x*y/1000,  just for the layout -- the numbers and BKG are aligned by hand, we can be more precise on aligment.  I also add how a exclusion line could appear (here is z-axis$<$20). } \label{fig:example} }
% \end{center}\end{figure*} 

To interpret the results with the benchmarks of Tab.~\ref{tab:BPs} we saturate the values of the VLQ couplings to obtain the needed size of the width for each point in the $(M_Q, \Gamma_Q / M_Q)$ grid of our scan. The total cross-section obtained by saturation is then compared to the experimental limits and it is verified to be excluded or not. In fact, the coupling necessary to fix $\Gamma_Q / M_Q$ have an effect on the cross-section of the process and therefore on the region of exclusion on the $(M_Q, \Gamma_Q / M_Q)$ plane, using the same data. 
% For example for a top partner VLQ $T$ we have:
% \begin{eqnarray}   
% \Gamma_{100\%} & = & \Gamma_{100\%}^0 (T \rightarrow W^+ q), \\
% \Gamma_{\text{singlet}} & = & \Gamma_{\text{singlet}}^0 (T \rightarrow W^+ q) + \Gamma_{\text{singlet}}^0 (T \rightarrow Z q^\prime) + \Gamma_{\text{singlet}}^0 (T \rightarrow H q^\prime).
% \end{eqnarray}
For instance, trivially, the value of the VLQ couplings will be larger for benchmark points where they interact 100\% in a specific channel than for the singlet- or doublet-like benchmarks, given the contribution from other decay channel(s) in the latter case.  

\begin{figure}[ht!]
\centering
\epsfig{file=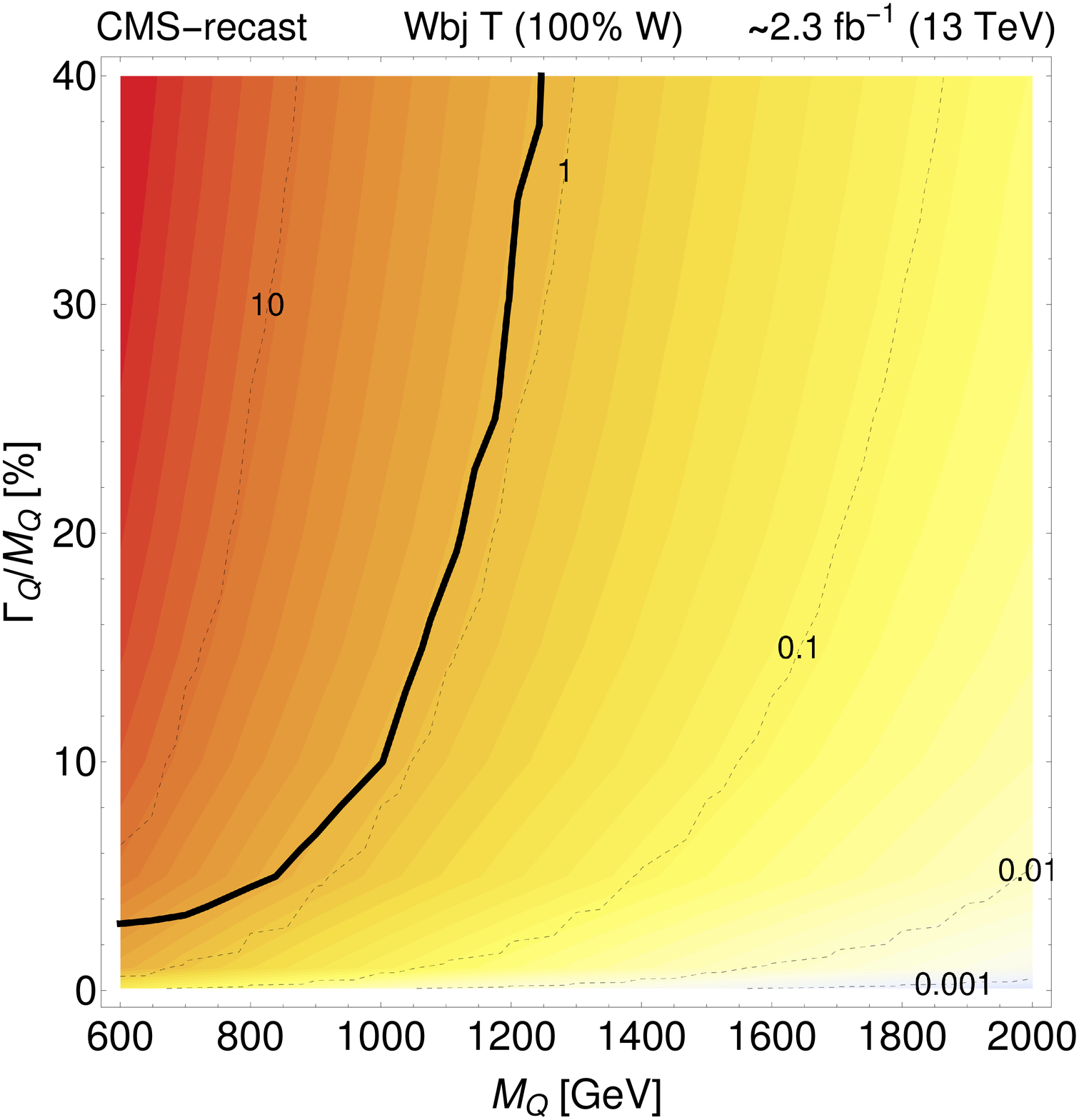,width=.45\textwidth}
\epsfig{file=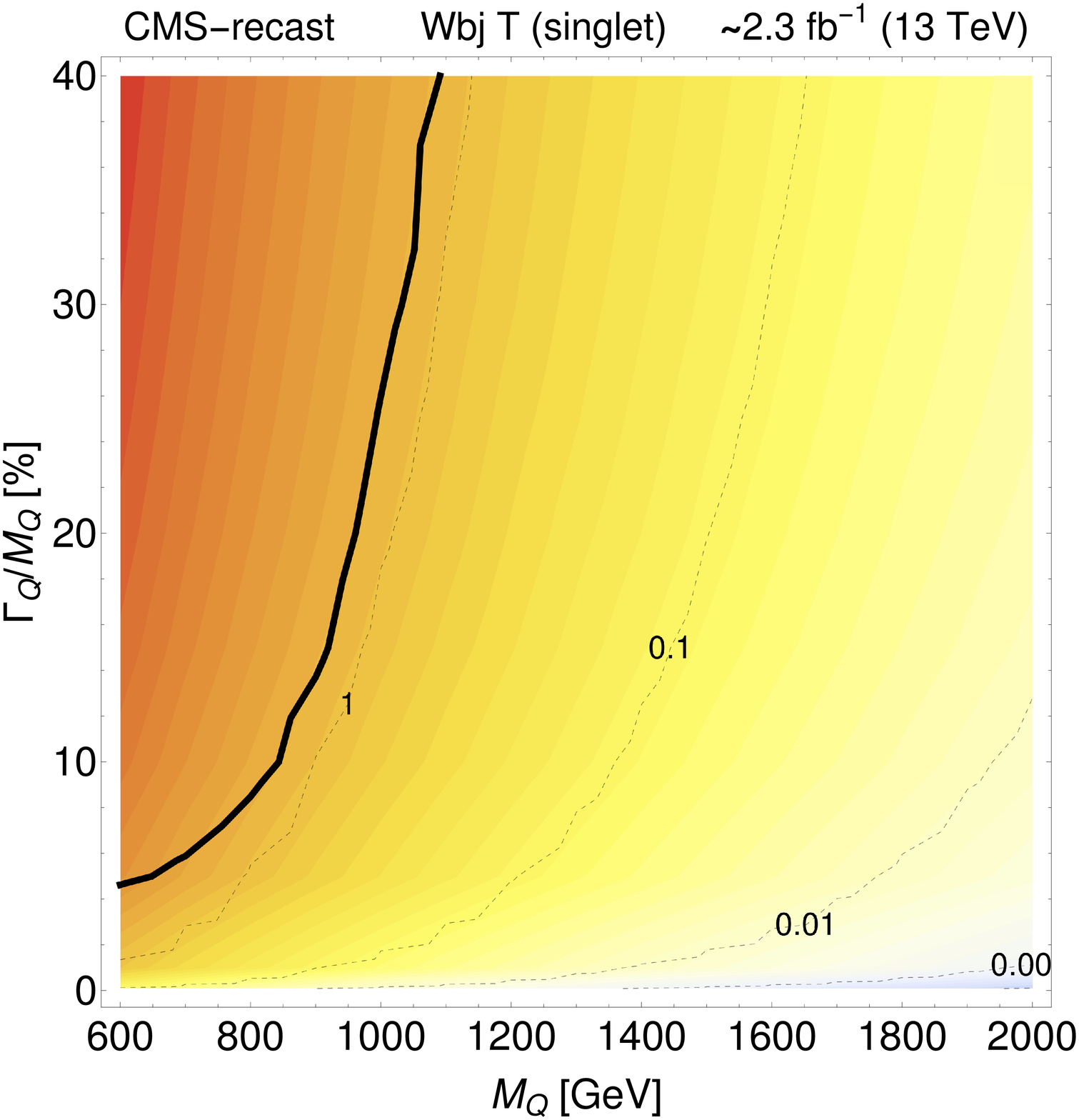,width=.45\textwidth}
\caption{\label{fig:combined plot} Cross section (in pb) for the $Wbj$ final state as a function of $M_T$ and of $\Gamma_T / M_T$ and 95\% exclusion line (in black) for a VLQ $T$ decaying 100\% to $bW$ (left panel) or singlet-like (right panel).}
\end{figure}

This done in  Fig. \ref{fig:combined plot}, where we show the value of the $Wbj$ signal cross-section as a function of the VLQ mass and of its $\Gamma_Q/M_Q$ ratio, together with the 95\% exclusion line for the $Wbj$ final state. It is possible to notice that the NWA region is not excluded while the exclusion range appears as $\Gamma_Q/M_Q$ increases, thus highlighting the phenomenological impact of the LW regime.  
The shape of the exclusion boundary is a balance between how the theory cross-section grows on the $(M_Q, \Gamma_Q / M_Q)$ plane and the behaviour of the signal kinematics. In Fig.~\ref{fig:combined plot} we see that the exclusion bounds roughly track the variation of the cross-section, with the exclusion line in the region where the cross-section is slightly larger than 1 pb. Such  behaviour is related to the fact that, for the considered CMS search, selection  cuts are designed in such a way that the signal efficiencies vary only moderately in the whole plane. In the light of this result, in the following, we will only show results for VLQ coupling which are 100\% to a specific SM boson, the results for singlet- and doublet-like cases being qualitatively analogous to a simple rescaling of the cross-section. 

% For these simulations we considered all the processes $p p \rightarrow W^+ \, b \, j$ and $p p \rightarrow W^- \, \bar b \, j$ containing exactly two VLQ couplings.

\begin{figure}[ht!]
\centering
\epsfig{file=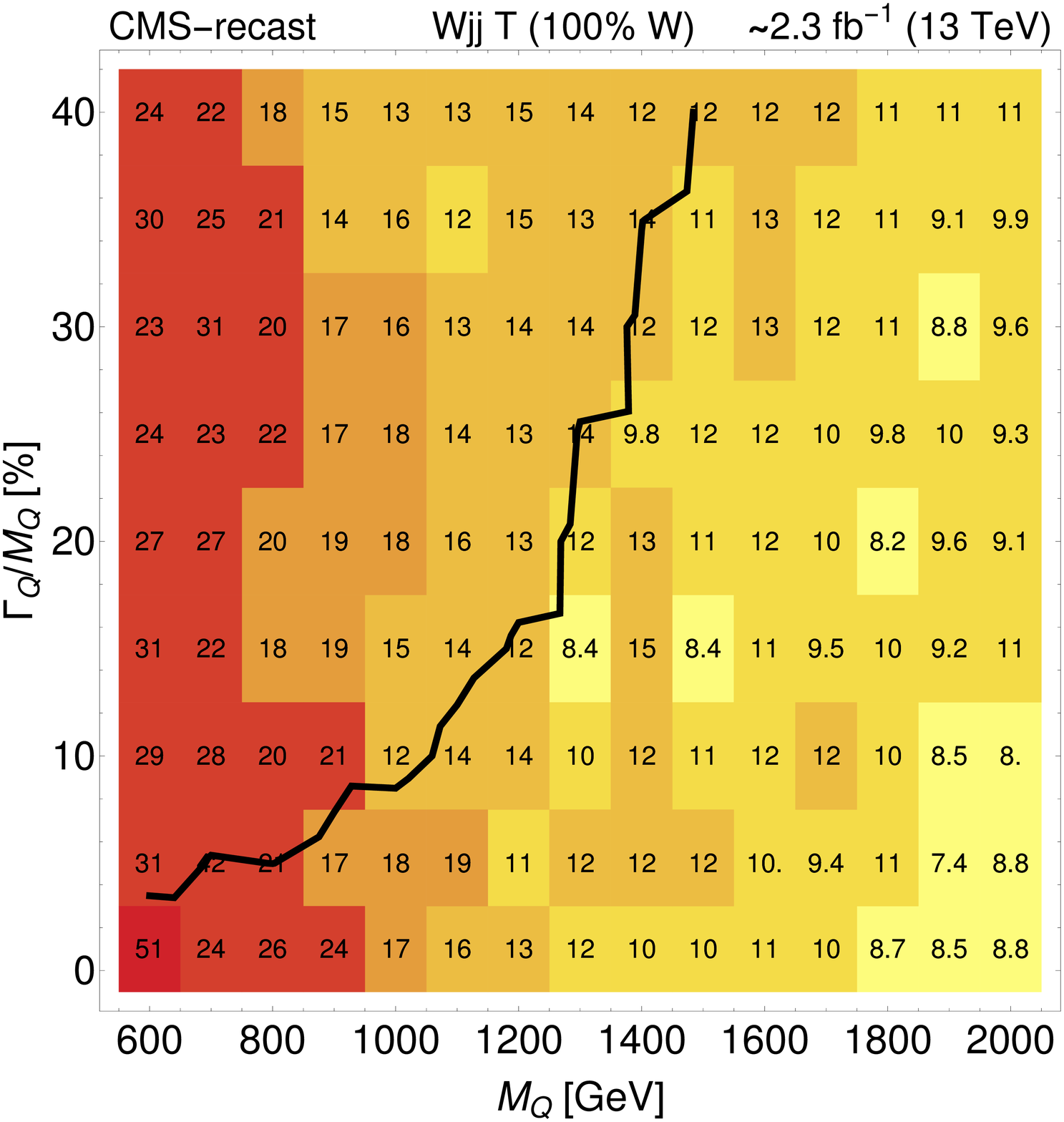,width=.45\textwidth}\\
\epsfig{file=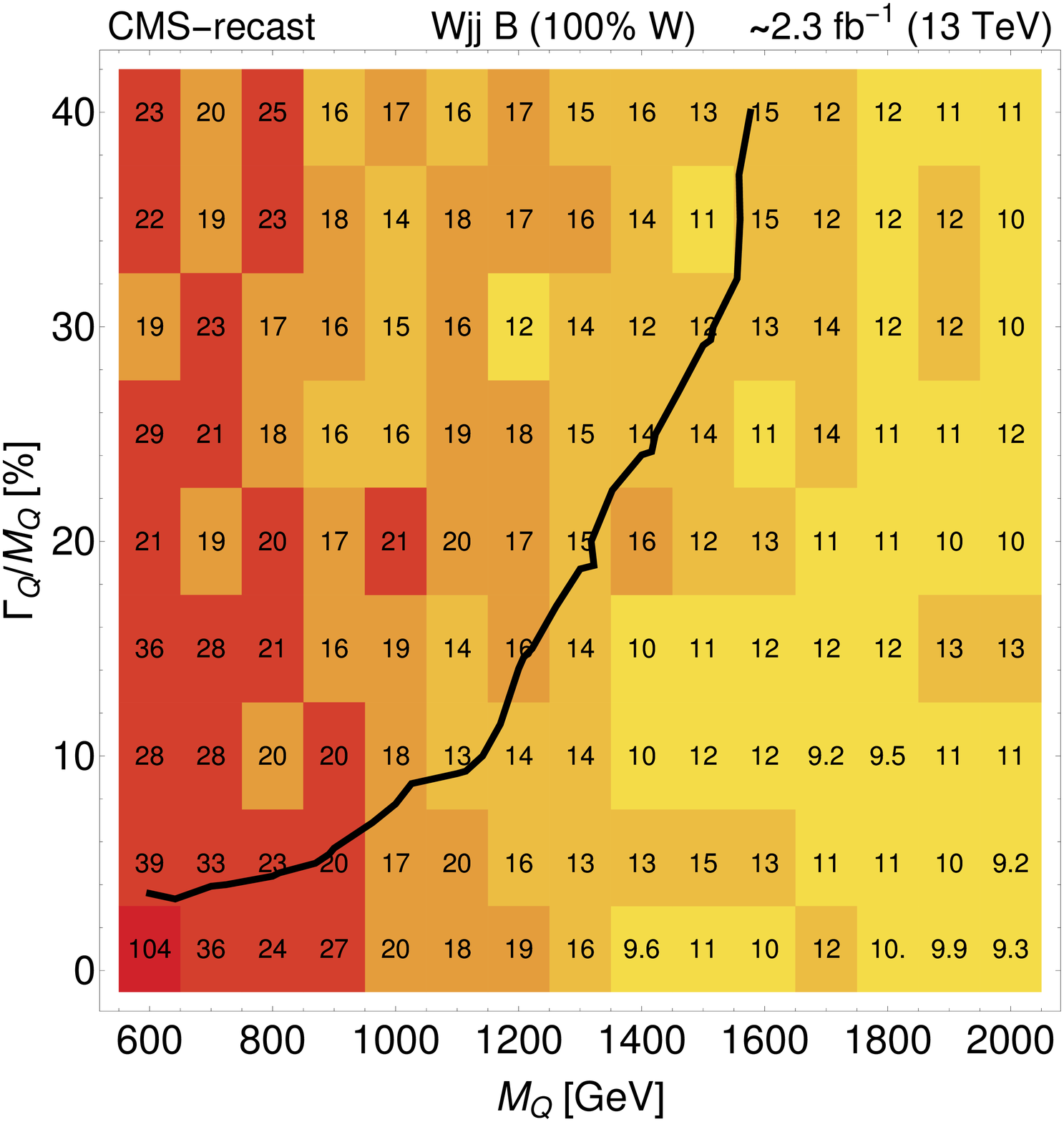,width=.45\textwidth}
\epsfig{file=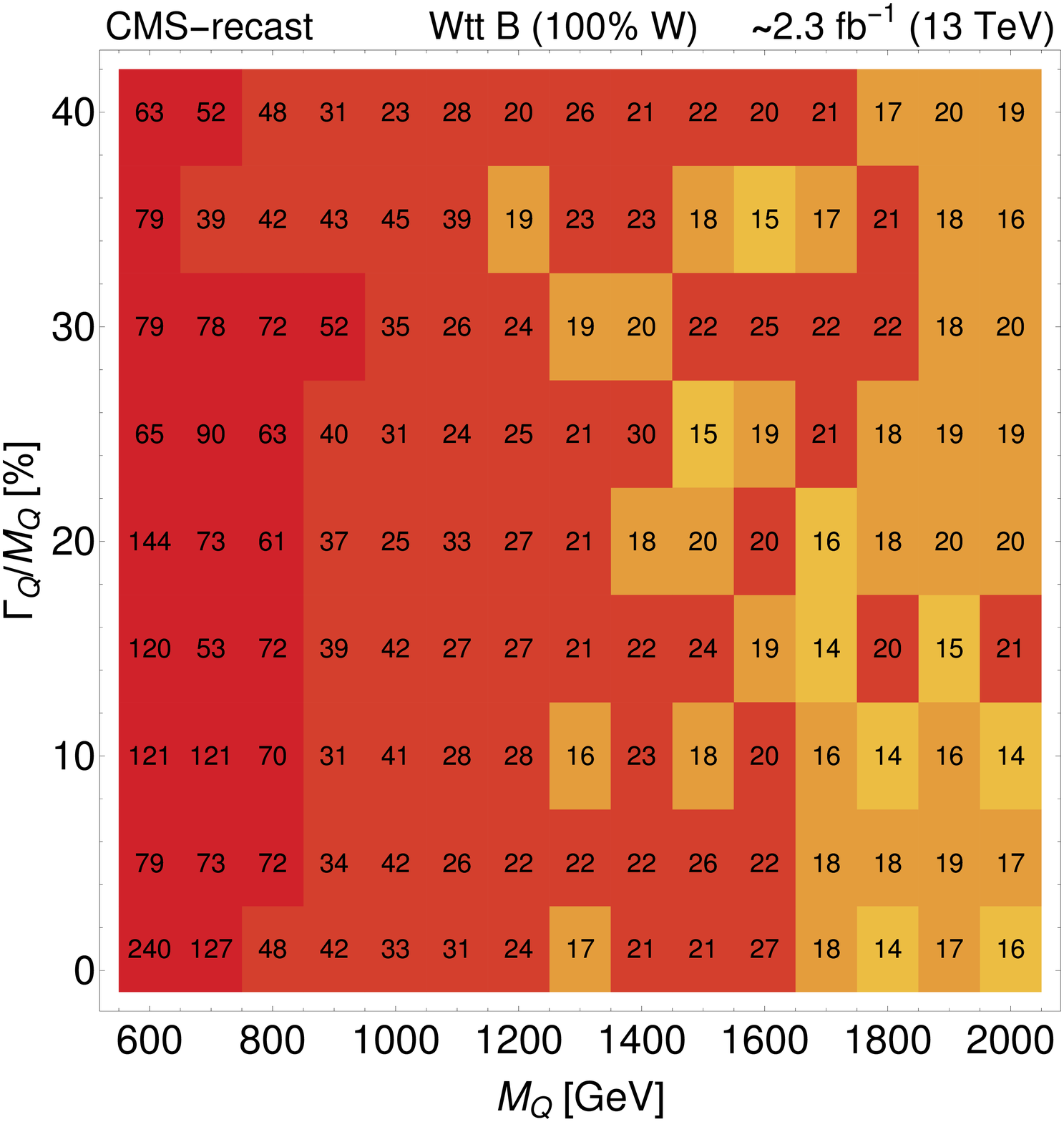,width=.45\textwidth}
\caption{\label{fig:sigma95rest} 95\% CL excluded cross-sections (in pb) and exclusion bound (black line) as a function of $M_Q$ and $\Gamma_Q / M_Q$ for a VLQ $T$ (top row) and $B$ (bottom row) coupling to the first (top and bottom-left panels) or third (bottom-right panel) SM quark generation.} 
\end{figure}

In Fig.~\ref{fig:sigma95rest} we show the values of the the largest allowed cross-section at 95\% CL, together with the exclusion bound, for each point of the grid in the $(M,\Gamma_Q/M_Q)$ plane for a VLQ $T$ (top row) or $B$ (bottom row) coupling to the first (top and bottom-left panels) or third (bottom-right panel) SM quark generations. 
% Here again we considered all the processes $p p \rightarrow W^+ \, q \, q$ and $p p \rightarrow W^- \, \bar q \, q$ (where $q$ can be a $b$, a $t$ or a jet) containing exactly two VLQ couplings. 
For the scenario where the $B$ couples with the third SM quark generation (bottom-right) the value of the the largest allowed cross-section is rather large in comparison with the other benchmarks we consider and therefore no exclusion can be made\footnote{This can be explained by a combination of two factors: the final state $Wtt$ is heavier than the other ones we consider, leading to a much smaller cross-section and,  since the search is not targeting this specific final state, the signal efficiencies are smaller than for the other channels.}. 
In contrast, we do find exclusion regions for the $Wjj$ channel mediated by a light generation $T$ or $B$ partner.

Analogously to the $Wbj$ case, the exclusion lines in both the $Wjj$ scenarios basically follow the increase of the cross-section due to the increase of the coupling. However, as the search is not optimised for  $Wjj$  final states, the largest allowed cross-section is much larger than for the $Wbj$ 
case\footnote{For instance, the search we are recasting requires a $b$ quark in the final state, which can only come from the accompanying jet when the VLQ only decays to light quarks.} The fact that we still get a similar exclusion for these final states is due to the higher cross-section for the light generation VLQs: PDF effects makes the $b$-initiated processes less likely that the $u$- and $d$-initiated ones, leading to a much larger cross-section for the $Wjj$ final states. Between the two $Wjj$ plot results, we observe a stronger exclusion in the case of the propagation of a $B$ VLQ, rather than a $T$ VLQ,  due to the fact that the 
cross-section is larger while the efficiencies are similar in the whole parameter space.

In all the channels we consider the largest allowed cross-section decreases when the mass increases, but, except for small masses (less than 1 TeV), this search is not really sensitive to the width variation. The effect of the width on the excluded cross-section is larger for $M_Q \simeq 600$ GeV where it can increase by a factor 4 with $\Gamma_Q / M_Q$ increasing from the NWA to $ \simeq 40$\%. The overall weak dependence of the results on the VLQ width is related to the fact that the kinematic distributions of the final states are also often weakly affected. The same weak dependence was found in the two CMS single production searches exploring the large VLQ width regime~\cite{Sirunyan:2017ynj,Sirunyan:2018fjh}.

%%%%%%%%%%%%%%%%%%%%%%%%%%%%%%%%%%%%%%%%%%%%%%%%%%%%%%%%%%%%%%%%%%%%%%%%%
\subsection{Reinterpretation through the reduced cross-section $\hat \sigma$}
\label{sec:reinterpretation}

In this section we provide an example of how to interpret experimental data only using the reduced cross-section $\hat\sigma$ described in Sect.~\ref{sec:parameterisation} for a model featuring a VLQ interaction with SM bosons with given couplings, mass and total width. The procedure can be generalised for any search of BSM states which can decay into SM final states, for which the width of the resonance can be considered as a further degree of freedom, complementary to the mass of the BSM state, for the determination of the maximum excluded cross-section. 

If interference terms, both between signal topologies and between signal and SM background are negligible, the cross-section for any process $p p \rightarrow V \; q \; q$ with $V$ any SM boson and $q$ any SM quark can be obtained %if the values of $\hat{\sigma}$ are provided together with the experimentally excluded cross-section, as it has been done in Refs.~\cite{Sirunyan:2017ynj,Sirunyan:2018fjh}. 
% For this purpose we provide a set of $\hat\sigma$ plots corresponding to any final state compatible with VLQ single production processes in \cite{} \Red{here we will put the link to the Phenodata page}, including the interference terms described in \eqref{eq:FWinterference}. 
by comparing the $\hat\sigma$ cross-sections, rescaled with appropriate couplings, to the excluded cross-sections (minding the model-dependent relation between the couplings and $\Gamma_Q / M_Q$) and determining whether a specific scenario is excluded by the experimental search without the need of performing a dedicated recast (as it was done to construct Fig.~\ref{fig:sigma95}). This is possible provided the results of the searches are presented in the $(M_Q, \Gamma_Q / M_Q)$ plane, as it has been done in the last section and in Refs.~\cite{Sirunyan:2017ynj,Sirunyan:2018fjh}.

\begin{figure}[ht!]
\centering
\epsfig{file=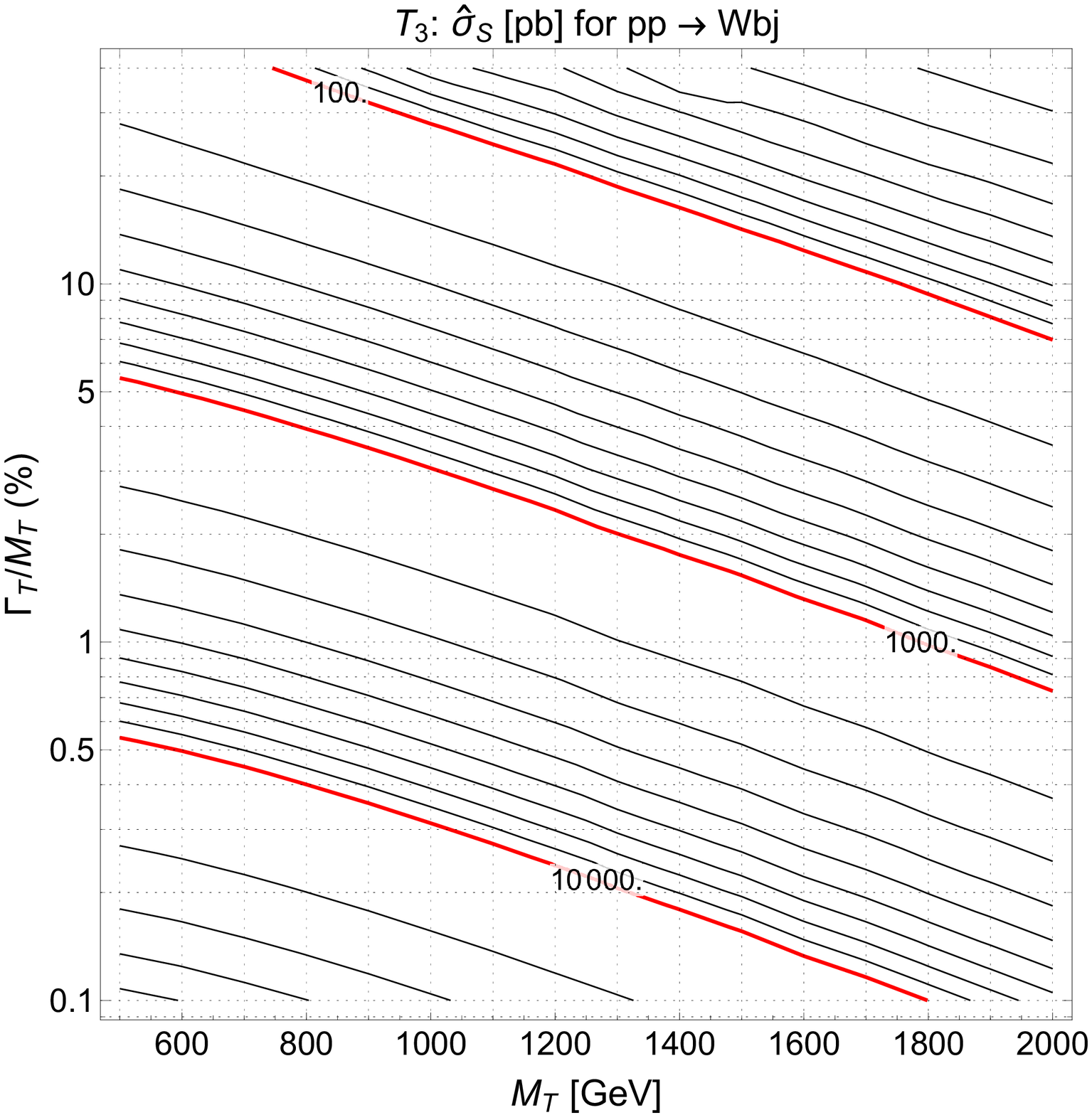,width=.45\textwidth}
\epsfig{file=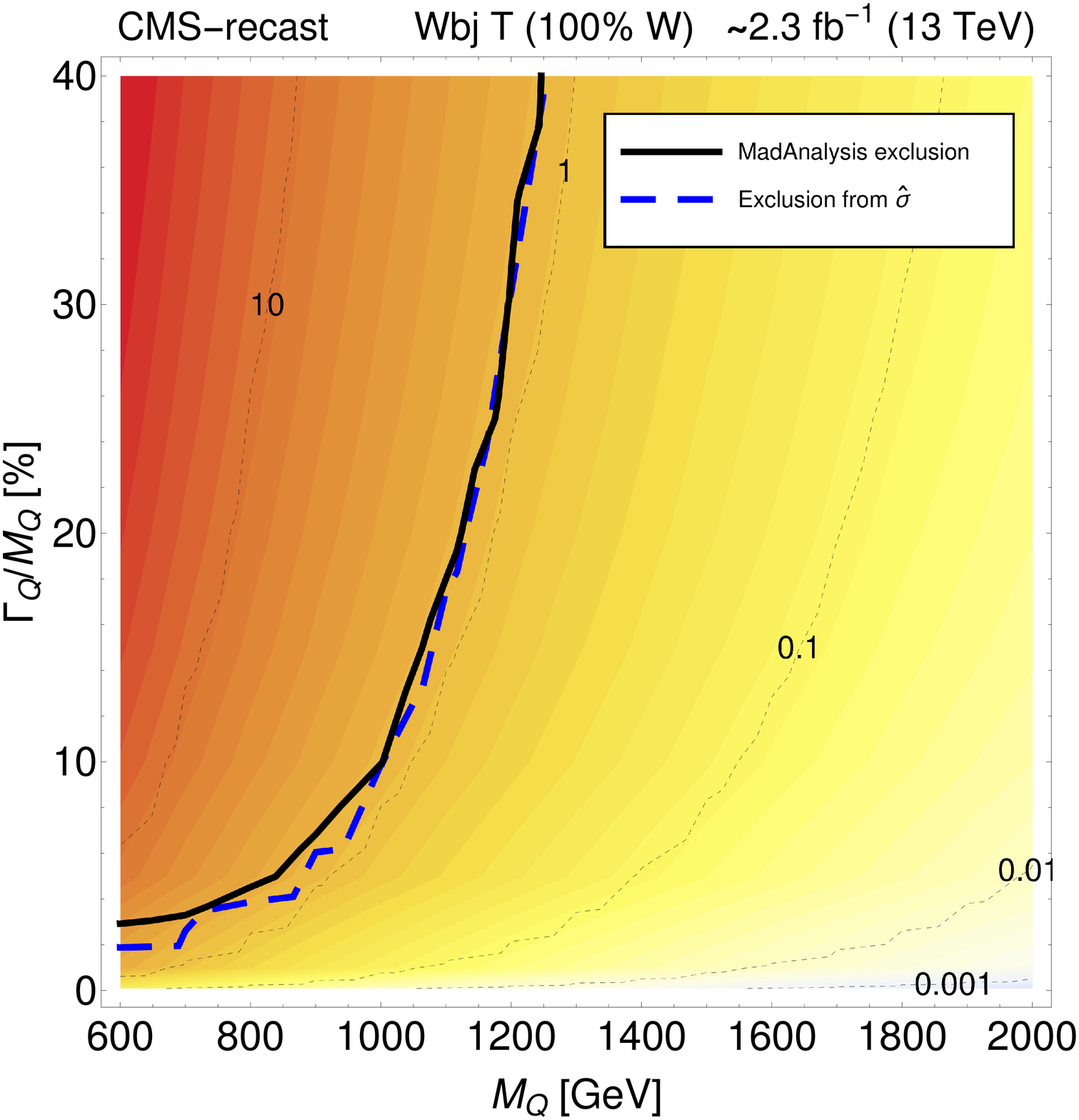,width=.45\textwidth}
\caption{\label{fig:rescaledSigmaHat} The 13 TeV $\hat{\sigma}$ values (left) and the signal cross-section (right), both given in pb, as a function of $M_Q$ and $\Gamma_Q / M_Q$ for a VLQ $T$ coupling to the third generation shown together with the exclusions line obtained using the $\hat{\sigma}$ data (in dashed blue) and with the original exclusion line (in black, same as Fig.~\ref{fig:combined plot}), assuming negligible interference contributions from any source.}
\end{figure}

In Fig.~\ref{fig:rescaledSigmaHat} the $\hat\sigma$'s for the process $pp\rightarrow bWj$ mediated by a $T$ interacting exclusively through charged current are shown on the left panel while on the right panel we report a comparison between the limit obtained through a numerical recasting and the limit obtained by rescaling the $\hat\sigma$ values with the coupling needed to obtain the corresponding width of the $T$ in this benchmark scenario. The results are qualitatively very close and allow to identify with good accuracy the excluded region of this benchmark scenario. 

\subsubsection{Including interference terms for the ${\rm BR}(T\to Wb)=100\%$ scenario}

Finally, we consider the role of interference between signal and background for the $pp\rightarrow bWj$ channel in the scenario with $T$ mixing with third SM quark generation and coupling only through charged current. Such assumptions allow us to study the scenario through the parameterisation of Eqs.~\eqref{eq:FW} and \eqref{eq:FWSMintlimit}. To perform the analysis, we start from the fact that, for any specific point, characterised by the mass and $\Gamma_Q/M_Q$ ratio of the VLQ, the total number of events is composed by different contributions:
\begin{equation}
\label{eq:S+B}
S+B = L (\sigma_S \epsilon_S + \sigma_{SB_{\rm{irr}}}^{\rm{int}} \epsilon_{SB_{\rm{irr}}}^{\rm{int}}) + B_{\rm{irr+red}} \equiv L \sigma_{\text{eff}} + B,
\end{equation}
where $L$ is the integrated luminosity, $B_{\rm{irr}}$ and $B_{\rm{red}}$ are the number of events for the irreducible and reducible background terms respectively (which sum to the total number of background events $B$) and $\sigma_{\text{eff}}\equiv \sigma_{S,\text{eff}} + \sigma_{SB_{\rm{irr}},\text{eff}}^{\rm{int}}$ is the fiducial cross-section that is accessed by the experiment. The latter can be described by folding the experimental efficiencies of each contribution in the expression of $\hat\sigma$ of Eqs.~\eqref{eq:FW} and \eqref{eq:FWSMintlimit} as:
\small\begin{eqnarray}
\label{eq:FWinterferencefoldedefficiencies}
\sigma_{\text{eff}}(C_2,M_Q,\Gamma_Q;\chi_Q) &=& C_2^4~\hat\sigma_S(M_Q,\Gamma_Q)~\epsilon_S(M_Q,\Gamma_Q) + C_2^2~\hat\sigma_{SB_{\rm{irr}}}^{\rm{int}}(M_Q,\Gamma_Q,\chi_Q)~\epsilon_{SB_{\rm{irr}}}^{\rm{int}}(M_Q,\Gamma_Q,\chi_Q)\nonumber\\
&\equiv& C_2^4~\hat\sigma_{S,\text{eff}}(M_Q,\Gamma_Q) + C_2^2~\hat\sigma_{SB_{\rm{irr}},\text{eff}}^{\rm{int}}(M_Q,\Gamma_Q,\chi_Q).
\end{eqnarray}

\begin{figure}[ht!]
\centering
\epsfig{file=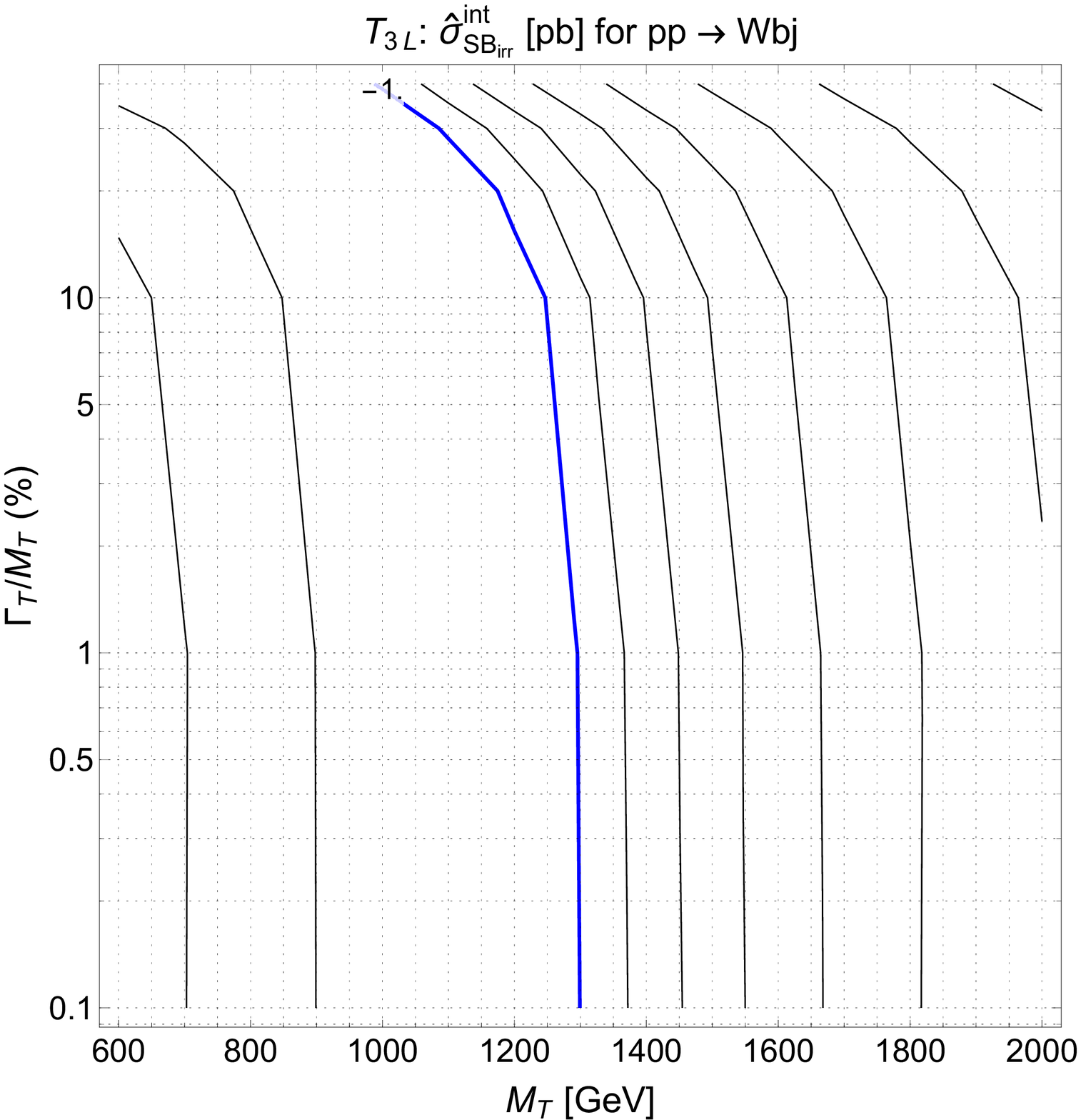,width=.45\textwidth}
\epsfig{file=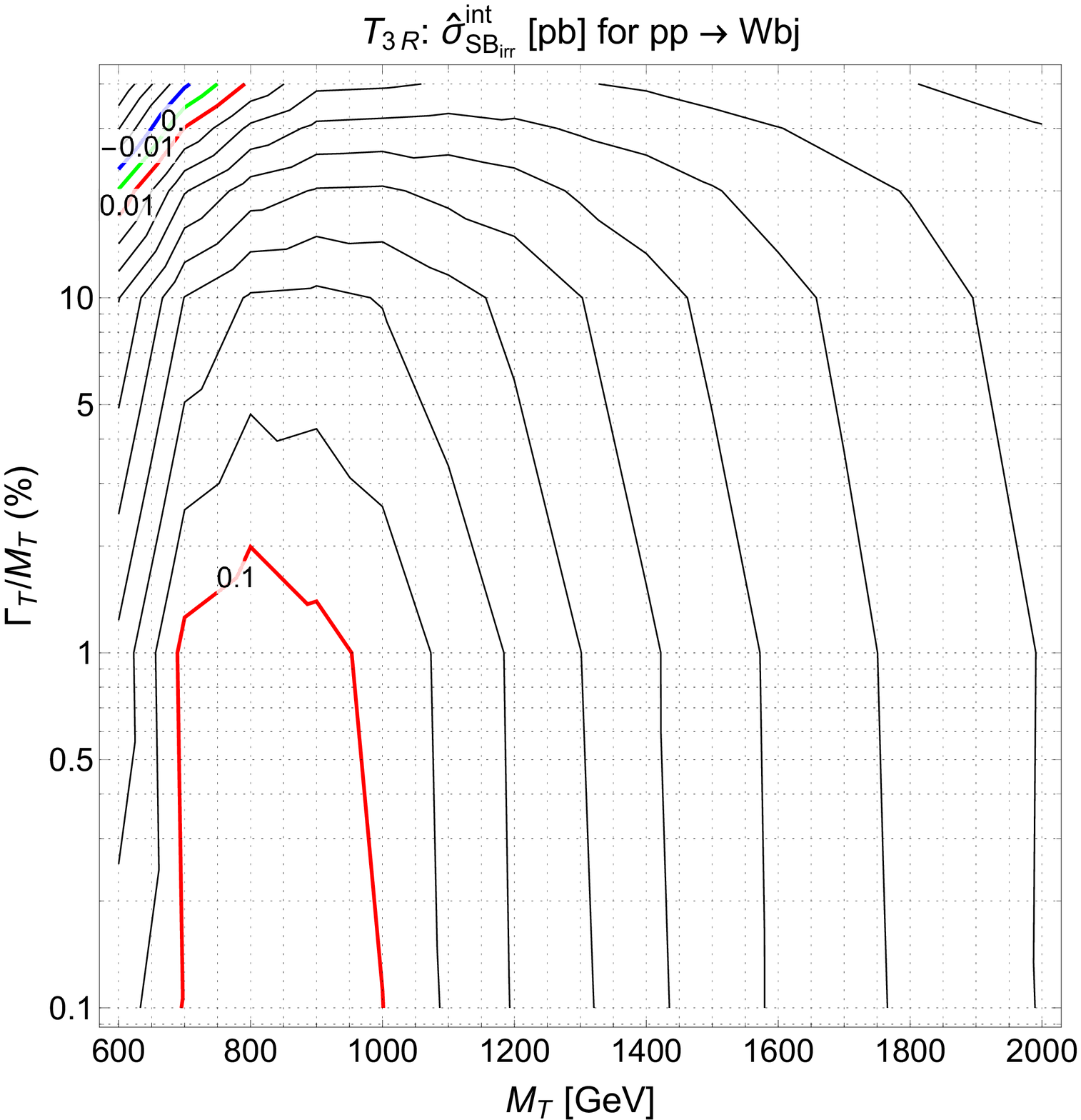,width=.45\textwidth}
\caption{\label{fig:sigmahatinterference} $\hat\sigma_{SB_{\rm{irr}}}^{\rm{int}}$ values at 13 TeV (in pb) for the process $pp\rightarrow bWj$ mediated by a $T$ interacting with third generation SM quarks exclusively through charged current, for the interference between signal and SM background with dominant LH (RH) coupling chirality of the $T$ in the left (right) panel.}
\end{figure}

The $\hat\sigma_{SB_{\rm{irr}}}^{\rm{int}}$ plots for the interference between signal and SM background are provided in Fig.~\ref{fig:sigmahatinterference}. Notice the different behaviour in the case of dominantly LH or RH coupling chirality of the $T$ and that, as expected, the scenario with dominant LH chirality produces larger interference contributions as the $T$ mass approaches the top mass region.

\begin{figure}[ht!]
\centering
\epsfig{file=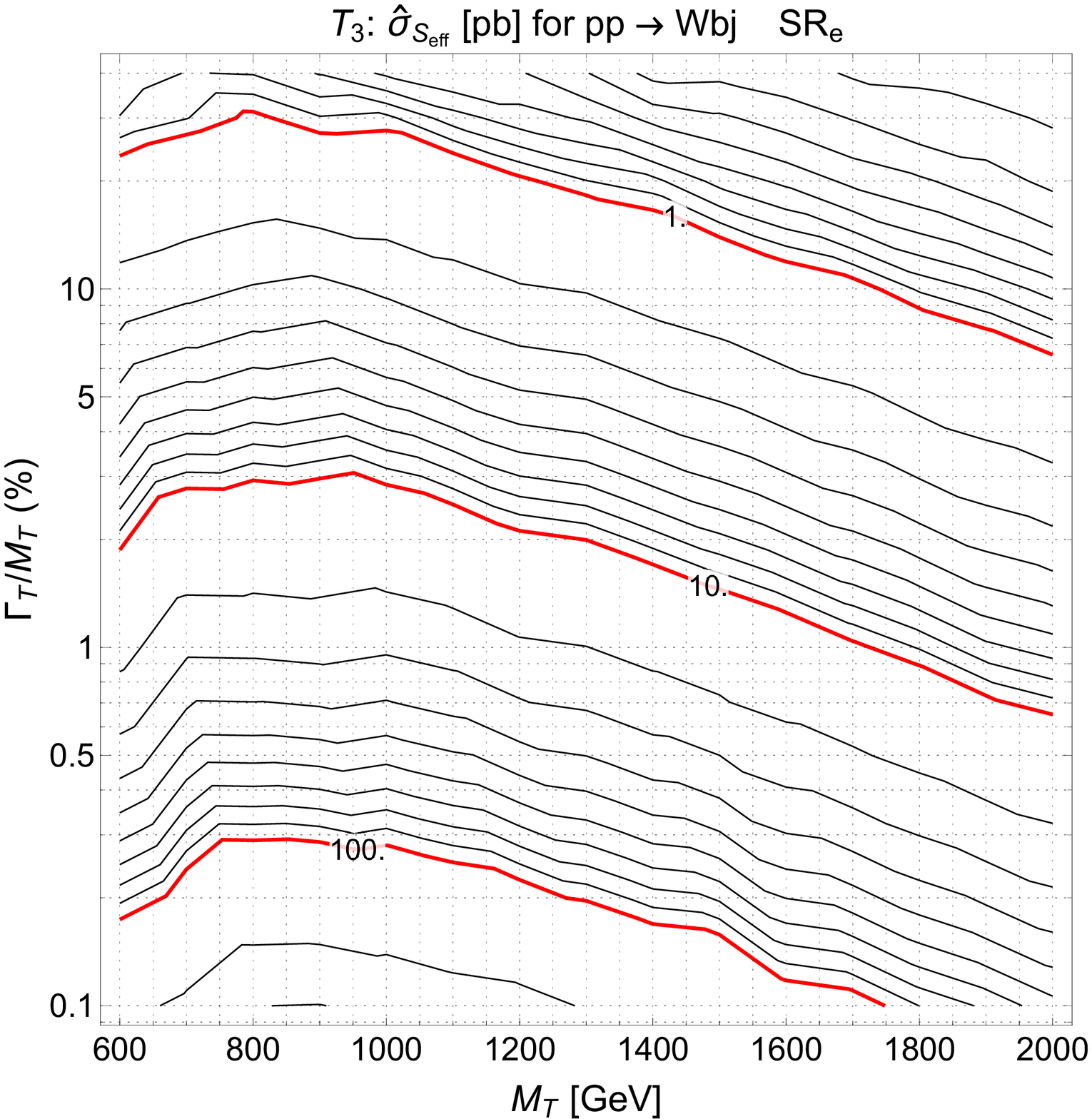,width=.45\textwidth}
\epsfig{file=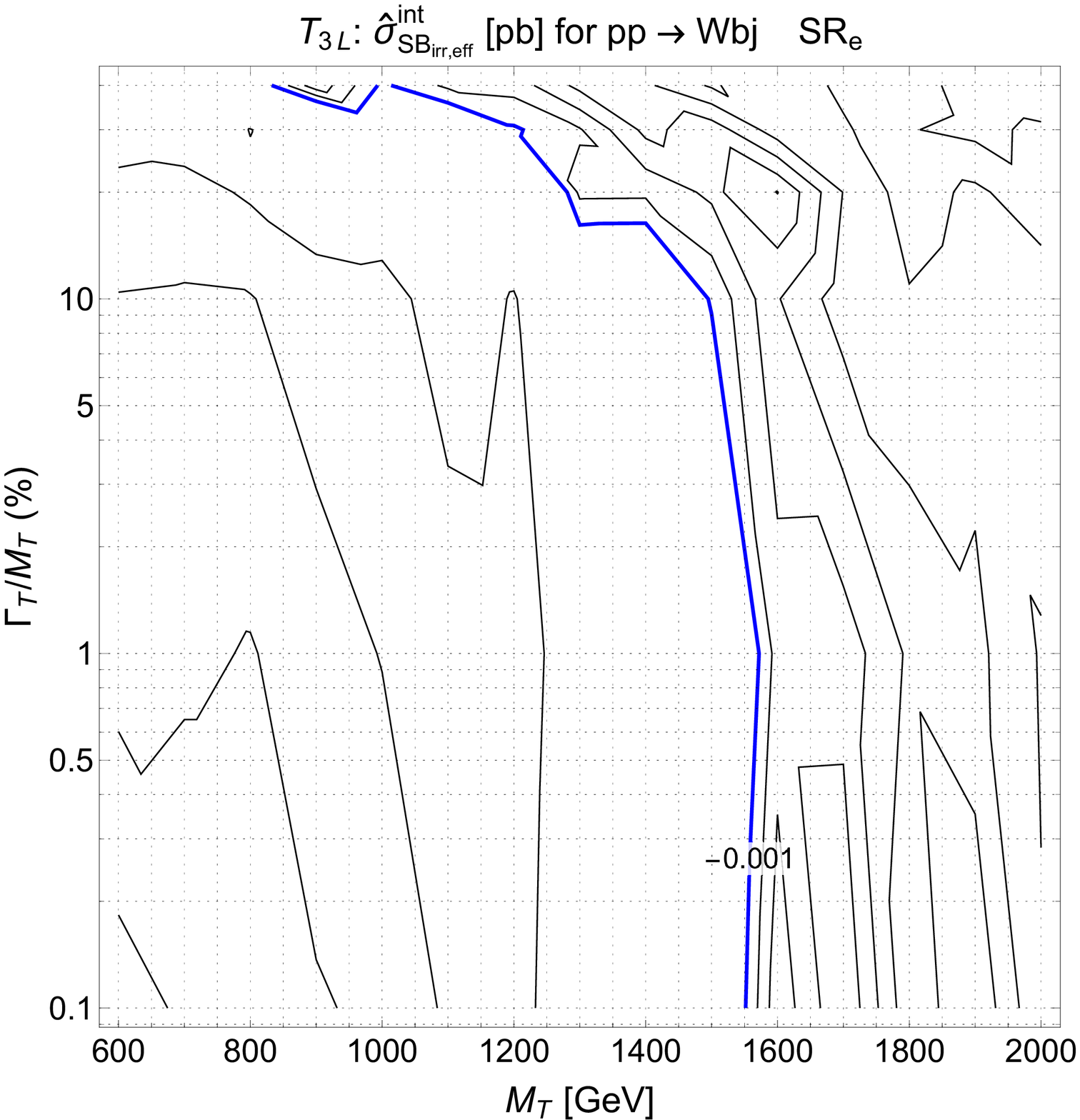,width=.45\textwidth}
\caption{\label{fig:sigmahateff} $\hat\sigma_{S,\text{eff}}$ and $\hat\sigma_{SB_{\rm{irr}},\text{eff}}^{\rm{int}}$ values at 13 TeV (in pb) for the process $pp\rightarrow bWj$ mediated by a $T$ interacting with third generation SM quarks exclusively through charged current, for the pure signal (left panel) and for the interference between signal and SM background with dominant LH coupling chirality of the $T$ (right panel). The signal region corresponding to the efficiencies is the single electron from the CMS search~\cite{Sirunyan:2017tfc}.}
\end{figure}

In Fig.~\ref{fig:sigmahateff} the values of $\hat\sigma_{S,\text{eff}}$ and $\hat\sigma_{SB_{\rm{irr}},\text{eff}}^{\rm{int}}$ are provided for the signal and for the interference term in the case of dominant LH chirality, considering the signal region targeting final states with an electron in the CMS search~\cite{Sirunyan:2017tfc}. With such information and with the knowledge of the total background $B$ (from experimental data), it is possible to reconstruct the number of events for $S+B$ for each point in the $(M_Q, \Gamma_Q / M_Q)$ plane through Eqs.~\eqref{eq:S+B} and \eqref{eq:FWinterferencefoldedefficiencies}. \\

\begin{figure}[ht!]
\centering
\epsfig{file=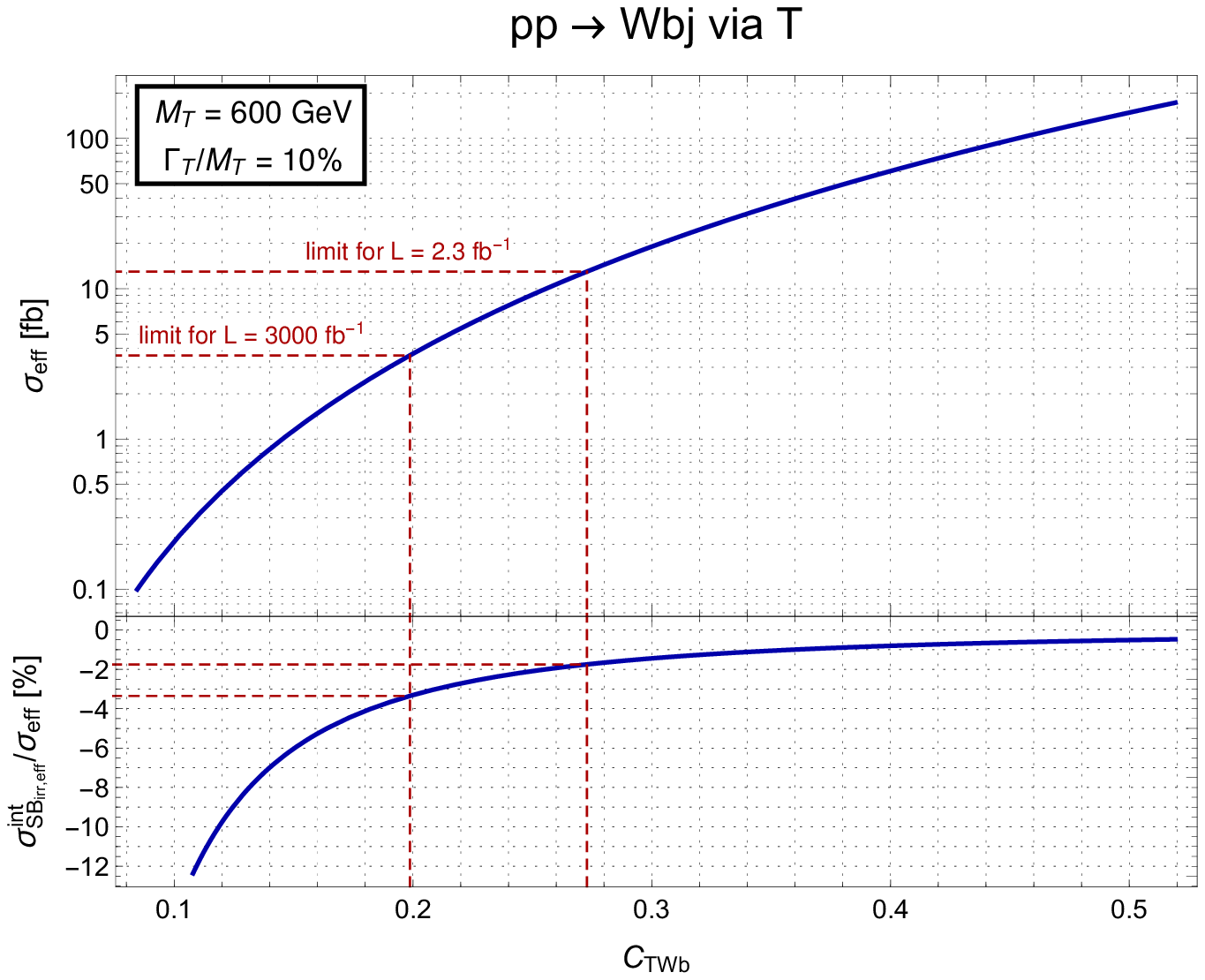,width=.7\textwidth}
\caption{\label{fig:interferencecoupling} Values of $\sigma_{\rm{eff}}$ (upper panel) and of the relative contribution of the interference term to it (lower panel) at 13 TeV as function of the $C_{TWb}$ coupling for the process $pp\rightarrow bWj$ mediated by a $T$ with $M_T=600$ GeV and $\Gamma_T/M_T=10$\% interacting with third generation SM quarks exclusively through charged current. Limits at two different LHC luminosities are also shown (see text for details).}
\end{figure}

The previous results for a $T$ decaying only through charged currents can be reinterpreted in terms of scenarios where other decay channels (which do not play a role as interference) are open, such as, e.g., $T \rightarrow S t$ for a (generic) new BSM scalar $S$. This allows us to modify the value of the coupling while keeping the $\Gamma_Q/M_Q$  ratio fixed\footnote{One should still check that the partial width $\Gamma (T \rightarrow W^+ b)$ is still smaller than the total width.}.
As a practical example of the procedure described above, we consider the point with $M_T=600$ GeV and $\Gamma_T/M_T=10\%$. The values of $\sigma_{\text{eff}}$ and the relative contribution of the interference term  are plotted as function of the $C_2\equiv C_{TWb}$ coupling in Fig.~\ref{fig:interferencecoupling}, considering  $C_{TWb}$ in a range for which $\sigma_{\rm{eff}}$ is larger than $10^{-1}$ fb up to when the partial width becomes equivalent to 60 GeV, ({i.e.}, 10\% of the mass). 
As it can also be inferred from Eq.~\eqref{eq:FWinterferencefoldedefficiencies}, the contribution of the interference term becomes more and more relevant as the $C_2$ coupling decreases, eventually becoming dominant with respect to the pure signal, but the fiducial cross-section $\sigma_{\rm{eff}}$ decreases as well. 
It is now straightforward to apply a CL analysis to determine the significance for exclusion or non-exclusion of the tested point. The reported numbers of background and observed events in the electron signal region by~\cite{Sirunyan:2017tfc} are $95\pm17$ and 78 respectively with a luminosity of 2.3 fb$^{-1}$, which imply that the quantity $\sigma_{\rm{eff}}*L\simeq30$ is excluded a 95\% CL, or that $\sigma_{\rm{eff}}$ has to be less than 13 fb. Such value places an upper bound on the $T$ coupling $C_{TWb}\lesssim0.27$, in a region for which the interference contribution to the signal is around -1.7\%. 

The relative importance of the interference contributions becomes larger as the luminosity increases. Rescaling the number of background events for the nominal highest luminosity for the LHC, 3 ab$^{-1}$, we obtain $B=105556$. With such value, and assuming a systematic uncertainty  for the background of 5\%, the maximum expected value of $\sigma_{\rm{eff}}$ for 95\% CL exclusion is 3.6 fb, corresponding to $C_{TWb}\lesssim0.2$, in a region for which the interference contribution to the signal is around -3.4\%. 

We stress that this example represents an ideal situation in which the signal and interference contributions can be factorised in a simple and gauge-invariant way. More complex scenarios, in which the VLQ can couple through neutral currents and/or other BSM states, in such a way that other interference terms are non-negligible, cannot be reinterpreted analogously to the example of this section.

%%%%%%%%%%%%%%%%%%%%%%%%%%%%%%%%%%%%%%%%%%%%%%%%%%%%%%%%%%%%%%%%%%%%%%%%%
%%%%%%%%%%%%%%%%%%%%%%%%%%%%%%%%%%%%%%%%%%%%%%%%%%%%%%%%%%%%%%%%%%%%%%%%%
\section{Conclusions}

We have studied in a model-independent way processes of single production of VLQs with non-negligible width at the LHC, exploiting a simple, yet robust, framework for the presentation of experimental results and for their subsequent reinterpretation in terms of any scenario of new physics beyond the SM which predict VLQs of generic masses, couplings and total width which can decay into SM final states. Such framework can be generalised also for different BSM scenarios or for analyses of single VLQ production where QCD NLO effects are considered.

We have assessed the role of the total width in the determination of kinematic distributions of the final state finding that, in some cases, the shapes of such distributions can significantly be deformed depending on the  $\Gamma_Q/M_Q$ ratio. 
We have numerically recast a CMS search to determine how bounds in the plane $(M_Q,\Gamma_Q/M_Q)$ change and, finally, we have compared the results of the numerical recasting in one of the considered scenarios to those from the model-independent procedure, obtaining an excellent agreement. Finally we have discussed the role of interference terms, usually neglected in experimental analysis, described issues related to the reinterpretation of results if such terms are not negligible and applied a procedure for reinterpretation in an ideal scenario in which such terms can be easily factorised.

The possibility to interpret observed data in terms of exclusion regions in the parameter space of models of new physics a is a crucial step for any phenomenological analysis of BSM scenarios. Nevertheless, in some cases, the experimental analysis techniques cannot be easily recast through phenomenological tools. In this case, for interpretations to be reliable, experimental results should provide all the essential information, so that a minimal, but complete, set of information can profitably be used for the characterisation of VLQs with generic assumptions on their mass, couplings and total width. In particular, the procedure adopted here allows for the possibility to perform a reinterpretation without the need of dedicated tools, thus growing the possibilities to test scenarios of new physics predicting VLQs.

%%%%%%%%%%%%%%%%%%%%%%%%%%%%%%%%%%%%%%%%%%%%%%%%%%%%%%%%%%%%%%%%%%%%%%%%%
%%%%%%%%%%%%%%%%%%%%%%%%%%%%%%%%%%%%%%%%%%%%%%%%%%%%%%%%%%%%%%%%%%%%%%%%%
\section*{Acknowledgements}

SM is funded in part through the NExT Institute and the STFC CG ST/P000711/1.

\bibliographystyle{JHEP}
\bibliography{XQCAT}

\providecommand{\href}[2]{#2}\begingroup\raggedright\begin{thebibliography}{10}

\bibitem{Aad:2012tfa}
{\scshape ATLAS} collaboration, G.~Aad et~al., \emph{{Observation of a new
  particle in the search for the Standard Model Higgs boson with the ATLAS
  detector at the LHC}},
  \href{https://doi.org/10.1016/j.physletb.2012.08.020}{\emph{Phys. Lett.}
  {\bfseries B716} (2012) 1} [\href{https://arxiv.org/abs/1207.7214}{{\ttfamily
  1207.7214}}].

\bibitem{Chatrchyan:2012xdj}
{\scshape CMS} collaboration, S.~Chatrchyan et~al., \emph{{Observation of a new
  boson at a mass of 125 GeV with the CMS experiment at the LHC}},
  \href{https://doi.org/10.1016/j.physletb.2012.08.021}{\emph{Phys. Lett.}
  {\bfseries B716} (2012) 30}
  [\href{https://arxiv.org/abs/1207.7235}{{\ttfamily 1207.7235}}].

\bibitem{Djouadi:2012ae}
A.~Djouadi and A.~Lenz, \emph{{Sealing the fate of a fourth generation of
  fermions}}, \href{https://doi.org/10.1016/j.physletb.2012.07.060}{\emph{Phys.
  Lett.} {\bfseries B715} (2012) 310}
  [\href{https://arxiv.org/abs/1204.1252}{{\ttfamily 1204.1252}}].

\bibitem{Eberhardt:2012gv}
O.~Eberhardt, G.~Herbert, H.~Lacker, A.~Lenz, A.~Menzel, U.~Nierste et~al.,
  \emph{{Impact of a Higgs boson at a mass of 126 GeV on the standard model
  with three and four fermion generations}},
  \href{https://doi.org/10.1103/PhysRevLett.109.241802}{\emph{Phys. Rev. Lett.}
  {\bfseries 109} (2012) 241802}
  [\href{https://arxiv.org/abs/1209.1101}{{\ttfamily 1209.1101}}].

\bibitem{Alves:2013dga}
A.~Alves, E.~Ramirez~Barreto, D.~Camargo and A.~Dias, \emph{{A Model with
  Chiral Quarks of Electric Charges -4/3 and 5/3}},
  \href{https://doi.org/10.1007/JHEP07(2013)129}{\emph{JHEP} {\bfseries 1307}
  (2013) 129} [\href{https://arxiv.org/abs/1306.1275}{{\ttfamily 1306.1275}}].

\bibitem{Banerjee:2013hxa}
S.~Banerjee, M.~Frank and S.~K. Rai, \emph{{Higgs data confronts Sequential
  Fourth Generation Fermions in the Higgs Triplet Model}},
  \href{https://doi.org/10.1103/PhysRevD.89.075005}{\emph{Phys. Rev.}
  {\bfseries D89} (2014) 075005}
  [\href{https://arxiv.org/abs/1312.4249}{{\ttfamily 1312.4249}}].

\bibitem{Holdom:2014bla}
B.~Holdom, \emph{{The accidental Higgs boson}},
  \href{https://doi.org/10.1103/PhysRevD.90.015004}{\emph{Phys. Rev.}
  {\bfseries D90} (2014) 015004}
  [\href{https://arxiv.org/abs/1404.6229}{{\ttfamily 1404.6229}}].

\bibitem{Bizot:2015zaa}
N.~Bizot and M.~Frigerio, \emph{{Fermionic extensions of the Standard Model in
  light of the Higgs couplings}},
  \href{https://doi.org/10.1007/JHEP01(2016)036}{\emph{JHEP} {\bfseries 01}
  (2016) 036} [\href{https://arxiv.org/abs/1508.01645}{{\ttfamily
  1508.01645}}].

\bibitem{Davidson:1987tr}
A.~Davidson and K.~C. Wali, \emph{{Family Mass Hierarchy From Universal Seesaw
  Mechanism}}, \href{https://doi.org/10.1103/PhysRevLett.60.1813}{\emph{Phys.
  Rev. Lett.} {\bfseries 60} (1988) 1813}.

\bibitem{Babu:1989rb}
K.~S. Babu and R.~N. Mohapatra, \emph{{A Solution to the Strong {CP} Problem
  Without an Axion}},
  \href{https://doi.org/10.1103/PhysRevD.41.1286}{\emph{Phys. Rev.} {\bfseries
  D41} (1990) 1286}.

\bibitem{Grinstein:2010ve}
B.~Grinstein, M.~Redi and G.~Villadoro, \emph{{Low Scale Flavor Gauge
  Symmetries}}, \href{https://doi.org/10.1007/JHEP11(2010)067}{\emph{JHEP}
  {\bfseries 11} (2010) 067} [\href{https://arxiv.org/abs/1009.2049}{{\ttfamily
  1009.2049}}].

\bibitem{Guadagnoli:2011id}
D.~Guadagnoli, R.~N. Mohapatra and I.~Sung, \emph{{Gauged Flavor Group with
  Left-Right Symmetry}},
  \href{https://doi.org/10.1007/JHEP04(2011)093}{\emph{JHEP} {\bfseries 04}
  (2011) 093} [\href{https://arxiv.org/abs/1103.4170}{{\ttfamily 1103.4170}}].

\bibitem{Moroi:1991mg}
T.~Moroi and Y.~Okada, \emph{{Radiative corrections to Higgs masses in the
  supersymmetric model with an extra family and antifamily}},
  \href{https://doi.org/10.1142/S0217732392000124}{\emph{Mod. Phys. Lett.}
  {\bfseries A7} (1992) 187}.

\bibitem{Moroi:1992zk}
T.~Moroi and Y.~Okada, \emph{{Upper bound of the lightest neutral Higgs mass in
  extended supersymmetric Standard Models}},
  \href{https://doi.org/10.1016/0370-2693(92)90091-H}{\emph{Phys. Lett.}
  {\bfseries B295} (1992) 73}.

\bibitem{Babu:2008ge}
K.~S. Babu, I.~Gogoladze, M.~U. Rehman and Q.~Shafi, \emph{{Higgs Boson Mass,
  Sparticle Spectrum and Little Hierarchy Problem in Extended MSSM}},
  \href{https://doi.org/10.1103/PhysRevD.78.055017}{\emph{Phys. Rev.}
  {\bfseries D78} (2008) 055017}
  [\href{https://arxiv.org/abs/0807.3055}{{\ttfamily 0807.3055}}].

\bibitem{Martin:2009bg}
S.~P. Martin, \emph{{Extra vector-like matter and the lightest Higgs scalar
  boson mass in low-energy supersymmetry}},
  \href{https://doi.org/10.1103/PhysRevD.81.035004}{\emph{Phys. Rev.}
  {\bfseries D81} (2010) 035004}
  [\href{https://arxiv.org/abs/0910.2732}{{\ttfamily 0910.2732}}].

\bibitem{Graham:2009gy}
P.~W. Graham, A.~Ismail, S.~Rajendran and P.~Saraswat, \emph{{A Little Solution
  to the Little Hierarchy Problem: A Vector-like Generation}},
  \href{https://doi.org/10.1103/PhysRevD.81.055016}{\emph{Phys. Rev.}
  {\bfseries D81} (2010) 055016}
  [\href{https://arxiv.org/abs/0910.3020}{{\ttfamily 0910.3020}}].

\bibitem{Martin:2010dc}
S.~P. Martin, \emph{{Raising the Higgs mass with Yukawa couplings for
  isotriplets in vector-like extensions of minimal supersymmetry}},
  \href{https://doi.org/10.1103/PhysRevD.82.055019}{\emph{Phys. Rev.}
  {\bfseries D82} (2010) 055019}
  [\href{https://arxiv.org/abs/1006.4186}{{\ttfamily 1006.4186}}].

\bibitem{Rosner:1985hx}
J.~L. Rosner, \emph{{E6 and Exotic Fermions}}, {\emph{Comments Nucl.Part.Phys.}
  {\bfseries 15} (1986) 195}.

\bibitem{Robinett:1985dz}
R.~Robinett, \emph{{On the Mixing and Production of Exotic Fermions in E6}},
  \href{https://doi.org/10.1103/PhysRevD.33.1908}{\emph{Phys.Rev.} {\bfseries
  D33} (1986) 1908}.

\bibitem{ArkaniHamed:2002qy}
N.~Arkani-Hamed, A.~Cohen, E.~Katz and A.~Nelson, \emph{{The Littlest Higgs}},
  \href{https://doi.org/10.1088/1126-6708/2002/07/034}{\emph{JHEP} {\bfseries
  0207} (2002) 034} [\href{https://arxiv.org/abs/hep-ph/0206021}{{\ttfamily
  hep-ph/0206021}}].

\bibitem{Schmaltz:2005ky}
M.~Schmaltz and D.~Tucker-Smith, \emph{{Little Higgs review}},
  \href{https://doi.org/10.1146/annurev.nucl.55.090704.151502}{\emph{Ann.Rev.Nucl.Part.Sci.}
  {\bfseries 55} (2005) 229}
  [\href{https://arxiv.org/abs/hep-ph/0502182}{{\ttfamily hep-ph/0502182}}].

\bibitem{Dobrescu:1997nm}
B.~A. Dobrescu and C.~T. Hill, \emph{{Electroweak symmetry breaking via top
  condensation seesaw}},
  \href{https://doi.org/10.1103/PhysRevLett.81.2634}{\emph{Phys.Rev.Lett.}
  {\bfseries 81} (1998) 2634}
  [\href{https://arxiv.org/abs/hep-ph/9712319}{{\ttfamily hep-ph/9712319}}].

\bibitem{Chivukula:1998wd}
R.~S. Chivukula, B.~A. Dobrescu, H.~Georgi and C.~T. Hill, \emph{{Top quark
  seesaw theory of electroweak symmetry breaking}},
  \href{https://doi.org/10.1103/PhysRevD.59.075003}{\emph{Phys.Rev.} {\bfseries
  D59} (1999) 075003} [\href{https://arxiv.org/abs/hep-ph/9809470}{{\ttfamily
  hep-ph/9809470}}].

\bibitem{He:2001fz}
H.-J. He, C.~T. Hill and T.~M. Tait, \emph{{Top quark seesaw, vacuum structure
  and electroweak precision constraints}},
  \href{https://doi.org/10.1103/PhysRevD.65.055006}{\emph{Phys.Rev.} {\bfseries
  D65} (2002) 055006} [\href{https://arxiv.org/abs/hep-ph/0108041}{{\ttfamily
  hep-ph/0108041}}].

\bibitem{Hill:2002ap}
C.~T. Hill and E.~H. Simmons, \emph{{Strong dynamics and electroweak symmetry
  breaking}},
  \href{https://doi.org/10.1016/S0370-1573(03)00140-6}{\emph{Phys.Rept.}
  {\bfseries 381} (2003) 235}
  [\href{https://arxiv.org/abs/hep-ph/0203079}{{\ttfamily hep-ph/0203079}}].

\bibitem{Agashe:2004rs}
K.~Agashe, R.~Contino and A.~Pomarol, \emph{{The Minimal composite Higgs
  model}},
  \href{https://doi.org/10.1016/j.nuclphysb.2005.04.035}{\emph{Nucl.Phys.}
  {\bfseries B719} (2005) 165}
  [\href{https://arxiv.org/abs/hep-ph/0412089}{{\ttfamily hep-ph/0412089}}].

\bibitem{Contino:2006qr}
R.~Contino, L.~Da~Rold and A.~Pomarol, \emph{{Light custodians in natural
  composite Higgs models}},
  \href{https://doi.org/10.1103/PhysRevD.75.055014}{\emph{Phys.Rev.} {\bfseries
  D75} (2007) 055014} [\href{https://arxiv.org/abs/hep-ph/0612048}{{\ttfamily
  hep-ph/0612048}}].

\bibitem{Barbieri:2007bh}
R.~Barbieri, B.~Bellazzini, V.~S. Rychkov and A.~Varagnolo, \emph{{The Higgs
  boson from an extended symmetry}},
  \href{https://doi.org/10.1103/PhysRevD.76.115008}{\emph{Phys.Rev.} {\bfseries
  D76} (2007) 115008} [\href{https://arxiv.org/abs/0706.0432}{{\ttfamily
  0706.0432}}].

\bibitem{Anastasiou:2009rv}
C.~Anastasiou, E.~Furlan and J.~Santiago, \emph{{Realistic Composite Higgs
  Models}}, \href{https://doi.org/10.1103/PhysRevD.79.075003}{\emph{Phys.Rev.}
  {\bfseries D79} (2009) 075003}
  [\href{https://arxiv.org/abs/0901.2117}{{\ttfamily 0901.2117}}].

\bibitem{AguilarSaavedra:2009es}
J.~Aguilar-Saavedra, \emph{{Identifying top partners at LHC}},
  \href{https://doi.org/10.1088/1126-6708/2009/11/030}{\emph{JHEP} {\bfseries
  0911} (2009) 030} [\href{https://arxiv.org/abs/0907.3155}{{\ttfamily
  0907.3155}}].

\bibitem{Okada:2012gy}
Y.~Okada and L.~Panizzi, \emph{{LHC signatures of vector-like quarks}},
  \href{https://doi.org/10.1155/2013/364936}{\emph{Adv.High Energy Phys.}
  {\bfseries 2013} (2013) 364936}
  [\href{https://arxiv.org/abs/1207.5607}{{\ttfamily 1207.5607}}].

\bibitem{DeSimone:2012fs}
A.~De~Simone, O.~Matsedonskyi, R.~Rattazzi and A.~Wulzer, \emph{{A First Top
  Partner Hunter's Guide}},
  \href{https://doi.org/10.1007/JHEP04(2013)004}{\emph{JHEP} {\bfseries 1304}
  (2013) 004} [\href{https://arxiv.org/abs/1211.5663}{{\ttfamily 1211.5663}}].

\bibitem{Buchkremer:2013bha}
M.~Buchkremer, G.~Cacciapaglia, A.~Deandrea and L.~Panizzi, \emph{{Model
  Independent Framework for Searches of Top Partners}},
  \href{https://doi.org/10.1016/j.nuclphysb.2013.08.010}{\emph{Nucl.Phys.}
  {\bfseries B876} (2013) 376}
  [\href{https://arxiv.org/abs/1305.4172}{{\ttfamily 1305.4172}}].

\bibitem{Aguilar-Saavedra:2013qpa}
J.~A. Aguilar-Saavedra, R.~Benbrik, S.~Heinemeyer and M.~Pérez-Victoria,
  \emph{{Handbook of vectorlike quarks: Mixing and single production}},
  \href{https://doi.org/10.1103/PhysRevD.88.094010}{\emph{Phys. Rev.}
  {\bfseries D88} (2013) 094010}
  [\href{https://arxiv.org/abs/1306.0572}{{\ttfamily 1306.0572}}].

\bibitem{Aaboud:2017qpr}
{\scshape ATLAS} collaboration, M.~Aaboud et~al., \emph{{Search for pair
  production of vector-like top quarks in events with one lepton, jets, and
  missing transverse momentum in $ \sqrt{s}=13 $ TeV $pp$ collisions with the
  ATLAS detector}}, \href{https://doi.org/10.1007/JHEP08(2017)052}{\emph{JHEP}
  {\bfseries 08} (2017) 052}
  [\href{https://arxiv.org/abs/1705.10751}{{\ttfamily 1705.10751}}].

\bibitem{Aaboud:2017zfn}
{\scshape ATLAS} collaboration, M.~Aaboud et~al., \emph{{Search for pair
  production of heavy vector-like quarks decaying to high-p$_{T}$ W bosons and
  b quarks in the lepton-plus-jets final state in pp collisions at $
  \sqrt{s}=13 $ TeV with the ATLAS detector}},
  \href{https://doi.org/10.1007/JHEP10(2017)141}{\emph{JHEP} {\bfseries 10}
  (2017) 141} [\href{https://arxiv.org/abs/1707.03347}{{\ttfamily
  1707.03347}}].

\bibitem{Aaboud:2018xuw}
{\scshape ATLAS} collaboration, M.~Aaboud et~al., \emph{{Search for pair
  production of up-type vector-like quarks and for four-top-quark events in
  final states with multiple $b$-jets with the ATLAS detector}},
  \href{https://arxiv.org/abs/1803.09678}{{\ttfamily 1803.09678}}.

\bibitem{Sirunyan:2017usq}
{\scshape CMS} collaboration, A.~M. Sirunyan et~al., \emph{{Search for pair
  production of vector-like T and B quarks in single-lepton final states using
  boosted jet substructure in proton-proton collisions at $\sqrt{s}=13$ TeV}},
  \href{https://doi.org/10.1007/JHEP11(2017)085}{\emph{JHEP} {\bfseries 11}
  (2017) 085} [\href{https://arxiv.org/abs/1706.03408}{{\ttfamily
  1706.03408}}].

\bibitem{Sirunyan:2017pks}
{\scshape CMS} collaboration, A.~M. Sirunyan et~al., \emph{{Search for pair
  production of vector-like quarks in the bW$\overline{\mathrm{b}}$W channel
  from proton-proton collisions at $\sqrt{s} =$ 13 TeV}},
  \href{https://doi.org/10.1016/j.physletb.2018.01.077}{\emph{Phys. Lett.}
  {\bfseries B779} (2018) 82}
  [\href{https://arxiv.org/abs/1710.01539}{{\ttfamily 1710.01539}}].

\bibitem{Sirunyan:2018omb}
{\scshape CMS} collaboration, A.~M. Sirunyan et~al., \emph{{Search for
  vector-like T and B quark pairs in final states with leptons at $\sqrt{s} =$
  13 TeV}},  \href{https://arxiv.org/abs/1805.04758}{{\ttfamily 1805.04758}}.

\bibitem{Sirunyan:2017ynj}
{\scshape CMS} collaboration, A.~M. Sirunyan et~al., \emph{{Search for single
  production of a vector-like T quark decaying to a Z boson and a top quark in
  proton-proton collisions at $\sqrt s$ = 13 TeV}},
  \href{https://doi.org/10.1016/j.physletb.2018.04.036}{\emph{Phys. Lett.}
  {\bfseries B781} (2018) 574}
  [\href{https://arxiv.org/abs/1708.01062}{{\ttfamily 1708.01062}}].

\bibitem{Sirunyan:2018fjh}
{\scshape CMS} collaboration, A.~M. Sirunyan et~al., \emph{{Search for single
  production of vector-like quarks decaying to a b quark and a Higgs boson}},
  \href{https://arxiv.org/abs/1802.01486}{{\ttfamily 1802.01486}}.

\bibitem{ATLAS:2016ovj}
{\scshape ATLAS} collaboration, \emph{{Search for single production of
  vector-like quarks decaying into $Wb$ in $pp$ collisions at $\sqrt{s} =$ 13
  TeV with the ATLAS detector}},  ATLAS-CONF-2016-072, 2016.

\bibitem{Barducci:2014ila}
D.~Barducci, A.~Belyaev, M.~Buchkremer, G.~Cacciapaglia, A.~Deandrea,
  S.~De~Curtis et~al., \emph{{Framework for Model Independent Analyses of
  Multiple Extra Quark Scenarios}},
  \href{https://doi.org/10.1007/JHEP12(2014)080}{\emph{JHEP} {\bfseries 12}
  (2014) 080} [\href{https://arxiv.org/abs/1405.0737}{{\ttfamily 1405.0737}}].

\bibitem{Barducci:2014gna}
D.~Barducci, A.~Belyaev, M.~Buchkremer, J.~Marrouche, S.~Moretti and
  L.~Panizzi, \emph{{XQCAT: eXtra Quark Combined Analysis Tool}},
  \href{https://doi.org/10.1016/j.cpc.2015.08.016}{\emph{Comput. Phys. Commun.}
  {\bfseries 197} (2015) 263}
  [\href{https://arxiv.org/abs/1409.3116}{{\ttfamily 1409.3116}}].

\bibitem{Barducci:2013zaa}
D.~Barducci, A.~Belyaev, J.~Blamey, S.~Moretti, L.~Panizzi and H.~Prager,
  \emph{{Towards model-independent approach to the analysis of interference
  effects in pair production of new heavy quarks}},
  \href{https://doi.org/10.1007/JHEP07(2014)142}{\emph{JHEP} {\bfseries 07}
  (2014) 142} [\href{https://arxiv.org/abs/1311.3977}{{\ttfamily 1311.3977}}].

\bibitem{Moretti:2016gkr}
S.~Moretti, D.~O'Brien, L.~Panizzi and H.~Prager, \emph{{Production of extra
  quarks at the Large Hadron Collider beyond the Narrow Width Approximation}},
  \href{https://doi.org/10.1103/PhysRevD.96.075035}{\emph{Phys. Rev.}
  {\bfseries D96} (2017) 075035}
  [\href{https://arxiv.org/abs/1603.09237}{{\ttfamily 1603.09237}}].

\bibitem{Moretti:2017qby}
S.~Moretti, D.~O'Brien, L.~Panizzi and H.~Prager, \emph{{Production of extra
  quarks decaying to Dark Matter beyond the Narrow Width Approximation at the
  LHC}}, \href{https://doi.org/10.1103/PhysRevD.96.035033}{\emph{Phys. Rev.}
  {\bfseries D96} (2017) 035033}
  [\href{https://arxiv.org/abs/1705.07675}{{\ttfamily 1705.07675}}].

\bibitem{Prager:2017owg}
H.~Prager, S.~Moretti, D.~O'Brien and L.~Panizzi, \emph{{Extra Quarks Decaying
  to Dark Matter Beyond the Narrow Width Approximation}},  in \emph{{5th Large
  Hadron Collider Physics Conference (LHCP 2017) Shanghai, China, May 15-20,
  2017}}, 2017, \href{https://arxiv.org/abs/1706.04001}{{\ttfamily
  1706.04001}},
  \href{http://inspirehep.net/record/1604905/files/arXiv:1706.04001.pdf}{http://inspirehep.net/record/1604905/files/arXiv:1706.04001.pdf}.

\bibitem{Prager:2017hnt}
H.~Prager, S.~Moretti, D.~O'Brien and L.~Panizzi, \emph{{Large width effects in
  processes of production of extra quarks decaying to Dark Matter at the LHC}},
   in \emph{{25th International Workshop on Deep-Inelastic Scattering and
  Related Topics (DIS 2017) Birmingham, UK, April 3-7, 2017}}, 2017,
  \href{https://arxiv.org/abs/1706.04007}{{\ttfamily 1706.04007}},
  \href{http://inspirehep.net/record/1604906/files/arXiv:1706.04007.pdf}{http://inspirehep.net/record/1604906/files/arXiv:1706.04007.pdf}.

\bibitem{Chen:2017hak}
C.-Y. Chen, S.~Dawson and E.~Furlan, \emph{{Vectorlike fermions and Higgs
  effective field theory revisited}},
  \href{https://doi.org/10.1103/PhysRevD.96.015006}{\emph{Phys. Rev.}
  {\bfseries D96} (2017) 015006}
  [\href{https://arxiv.org/abs/1703.06134}{{\ttfamily 1703.06134}}].

\bibitem{Campbell:2009ss}
J.~M. Campbell, R.~Frederix, F.~Maltoni and F.~Tramontano,
  \emph{{Next-to-Leading-Order Predictions for t-Channel Single-Top Production
  at Hadron Colliders}},
  \href{https://doi.org/10.1103/PhysRevLett.102.182003}{\emph{Phys. Rev. Lett.}
  {\bfseries 102} (2009) 182003}
  [\href{https://arxiv.org/abs/0903.0005}{{\ttfamily 0903.0005}}].

\bibitem{Frederix:2012dh}
R.~Frederix, E.~Re and P.~Torrielli, \emph{{Single-top t-channel
  hadroproduction in the four-flavour scheme with POWHEG and aMC@NLO}},
  \href{https://doi.org/10.1007/JHEP09(2012)130}{\emph{JHEP} {\bfseries 09}
  (2012) 130} [\href{https://arxiv.org/abs/1207.5391}{{\ttfamily 1207.5391}}].

\bibitem{Bothmann:2017jfv}
E.~Bothmann, F.~Krauss and M.~Schönherr, \emph{{Single top-quark production
  with SHERPA}},
  \href{https://doi.org/10.1140/epjc/s10052-018-5696-1}{\emph{Eur. Phys. J.}
  {\bfseries C78} (2018) 220}
  [\href{https://arxiv.org/abs/1711.02568}{{\ttfamily 1711.02568}}].

\bibitem{Brooijmans:2018xbu}
G.~Brooijmans et~al., \emph{{Les Houches 2017: Physics at TeV Colliders New
  Physics Working Group Report}},  in \emph{{10th Les Houches Workshop on
  Physics at TeV Colliders (PhysTeV 2017) Les Houches, France, June 5-23,
  2017}}, 2018, \href{https://arxiv.org/abs/1803.10379}{{\ttfamily
  1803.10379}},
  \href{https://inspirehep.net/record/1664565/files/1803.10379.pdf}{https://inspirehep.net/record/1664565/files/1803.10379.pdf}.

\bibitem{Alwall:2011uj}
J.~Alwall, M.~Herquet, F.~Maltoni, O.~Mattelaer and T.~Stelzer, \emph{{MadGraph
  5 : Going Beyond}},
  \href{https://doi.org/10.1007/JHEP06(2011)128}{\emph{JHEP} {\bfseries 1106}
  (2011) 128} [\href{https://arxiv.org/abs/1106.0522}{{\ttfamily 1106.0522}}].

\bibitem{Alwall:2014hca}
J.~Alwall, R.~Frederix, S.~Frixione, V.~Hirschi, F.~Maltoni, O.~Mattelaer
  et~al., \emph{{The automated computation of tree-level and next-to-leading
  order differential cross sections, and their matching to parton shower
  simulations}}, \href{https://doi.org/10.1007/JHEP07(2014)079}{\emph{JHEP}
  {\bfseries 07} (2014) 079} [\href{https://arxiv.org/abs/1405.0301}{{\ttfamily
  1405.0301}}].

\bibitem{Ball:2014uwa}
{\scshape NNPDF} collaboration, R.~D. Ball et~al., \emph{{Parton distributions
  for the LHC Run II}},
  \href{https://doi.org/10.1007/JHEP04(2015)040}{\emph{JHEP} {\bfseries 04}
  (2015) 040} [\href{https://arxiv.org/abs/1410.8849}{{\ttfamily 1410.8849}}].

\bibitem{Sjostrand:2007gs}
T.~Sjostrand, S.~Mrenna and P.~Z. Skands, \emph{{A Brief Introduction to PYTHIA
  8.1}}, \href{https://doi.org/10.1016/j.cpc.2008.01.036}{\emph{Comput. Phys.
  Commun.} {\bfseries 178} (2008) 852}
  [\href{https://arxiv.org/abs/0710.3820}{{\ttfamily 0710.3820}}].

\bibitem{Sjostrand:2006za}
T.~Sjostrand, S.~Mrenna and P.~Z. Skands, \emph{{PYTHIA 6.4 Physics and
  Manual}}, \href{https://doi.org/10.1088/1126-6708/2006/05/026}{\emph{JHEP}
  {\bfseries 0605} (2006) 026}
  [\href{https://arxiv.org/abs/hep-ph/0603175}{{\ttfamily hep-ph/0603175}}].

\bibitem{Conte:2014zja}
E.~Conte, B.~Dumont, B.~Fuks and C.~Wymant, \emph{{Designing and recasting LHC
  analyses with MadAnalysis 5}},
  \href{https://doi.org/10.1140/epjc/s10052-014-3103-0}{\emph{Eur.Phys.J.}
  {\bfseries C74} (2014) 3103}
  [\href{https://arxiv.org/abs/1405.3982}{{\ttfamily 1405.3982}}].

\bibitem{deFavereau:2013fsa}
{\scshape DELPHES 3} collaboration, J.~de~Favereau et~al., \emph{{DELPHES 3, A
  modular framework for fast simulation of a generic collider experiment}},
  \href{https://doi.org/10.1007/JHEP02(2014)057}{\emph{JHEP} {\bfseries 1402}
  (2014) 057} [\href{https://arxiv.org/abs/1307.6346}{{\ttfamily 1307.6346}}].

\bibitem{Alloul:2013bka}
A.~Alloul, N.~D. Christensen, C.~Degrande, C.~Duhr and B.~Fuks,
  \emph{{FeynRules 2.0 - A complete toolbox for tree-level phenomenology}},
  \href{https://doi.org/10.1016/j.cpc.2014.04.012}{\emph{Comput.Phys.Commun.}
  {\bfseries 185} (2014) 2250}
  [\href{https://arxiv.org/abs/1310.1921}{{\ttfamily 1310.1921}}].

\bibitem{Degrande:2011ua}
C.~Degrande, C.~Duhr, B.~Fuks, D.~Grellscheid, O.~Mattelaer et~al., \emph{{UFO
  - The Universal FeynRules Output}},
  \href{https://doi.org/10.1016/j.cpc.2012.01.022}{\emph{Comput.Phys.Commun.}
  {\bfseries 183} (2012) 1201}
  [\href{https://arxiv.org/abs/1108.2040}{{\ttfamily 1108.2040}}].

\bibitem{Fuks:2016ftf}
B.~Fuks and H.-S. Shao, \emph{{QCD next-to-leading-order predictions matched to
  parton showers for vector-like quark models}},
  \href{https://doi.org/10.1140/epjc/s10052-017-4686-z}{\emph{Eur. Phys. J.}
  {\bfseries C77} (2017) 135}
  [\href{https://arxiv.org/abs/1610.04622}{{\ttfamily 1610.04622}}].

\bibitem{Sirunyan:2017tfc}
{\scshape CMS} collaboration, A.~M. Sirunyan et~al., \emph{{Search for single
  production of vector-like quarks decaying into a b quark and a W boson in
  proton-proton collisions at $\sqrt s =$ 13 TeV}},
  \href{https://doi.org/10.1016/j.physletb.2017.07.022}{\emph{Phys. Lett.}
  {\bfseries B772} (2017) 634}
  [\href{https://arxiv.org/abs/1701.08328}{{\ttfamily 1701.08328}}].

\end{thebibliography}\endgroup

\end{document}